\documentclass[a4paper,10pt]{article}
\pdfoutput=1 
\usepackage{jcappub}
\usepackage[T1]{fontenc}
\usepackage{appendix}
\usepackage{enumitem}
\usepackage{booktabs}
\usepackage{multirow}
\usepackage{siunitx}
\usepackage{orcidlink}
\usepackage{arydshln}
\usepackage{caption}
\usepackage{subfig}
\usepackage{graphicx}

\title{\boldmath Inverse Cosmography: testing the effectiveness of cosmographic polynomials using machine learning}

\author[a]{Cristian Zamora Mun\~oz}
\author[a]{and Celia Escamilla-Rivera\orcidlink{0000-0002-8929-250X}}

\affiliation[a]{Instituto de Ciencias Nucleares, Universidad Nacional Aut\'onoma de M\'exico, Circuito Exterior C.U., A.P. 70-543, M\'exico D.F. 04510, M\'exico.}
\emailAdd{cristian.z.m@ciencias.unam.mx}
\emailAdd{celia.escamilla@nucleares.unam.mx}

\abstract
{Cosmography has been referred to as a solution to the \textit{inverse scattering problem}, which is reasonable since it allows us to calculate cosmological bounds from data samples by performing an expansion of the cosmological observables around the present time. Nevertheless, this approach is not \textit{properly} an inverse scattering solution since the method only circumvents the problem to fit the equation of state (EoS) parameters (model-dependent) by replacing it with their fit of its cosmographic parameters (with a polynomial series-dependence). Therefore, the question that we want to answer is: can we construct a \textit{new cosmography approach} where the cosmodynamical parameters can be fitted and then employ them to analyse the kinematics via its \textit{generic} cosmographic parameters? By all means, without experimenting with the traditional problem of truncation of the series that all cosmography proposals in the literature argue. In this work, we present a solution to this question.  A \textit generic EoS depending solely on the form of $f_{i}(z)$ and its derivative is found, where this function can be any polynomial (mimicking a dark energy-like term) that allow the dynamics of a specific cosmological density. We test our generic EoS with standard cosmological models and with polynomials proposals as Pad\'e and Chebyshev approximants. All of them reproduce $\Lambda$CDM at $z>1$ between 1-$\sigma$, but fail at lower observational redshift range. Interesting enough, a Pad\'e (2,2) approximant has been considered inside a $f(z)$CDM-like model showing a transition in $z=1$. Also, we found that this is not assured since models with these characteristics have degeneracy and truncation problems that have a divergence at this redshift limit. With a Chebyshev (2,1) approximant, here proposed, the divergence is not present for large redshifts. To explore our results, also we present a new supernovae sample trained via a deep learning tool called Recurrent-Bayesian (RNN+BNN) network that can solve problems as overfitting at lower redshifts and increase the density of data points in this region, which can help to discern between cosmographies at 2-$\sigma$ of precision. }

\begin{document}
\maketitle
\flushbottom


\section{Introduction}

To understand the cosmic accelerating evolution in a standard manner we often base our description of the Universe in a model-dependent way. This argument leads to a biased  cosmic reconstruction of the dynamics of the universe, where the cosmological parameters suffer from degeneracies problems or cannot be well-tested with a good-fit convergence using observational samples. Currently, these problems fall into the window of \textit{cosmological tension parameter discussions}, like those of $\sigma_8$ and $H_0$ \cite{Verde:2019ivm}. Strategies towards a model-independent analysis to relax these problems require  
the assumptions of homogeneity and isotropy in a scenario where the cosmic dynamics is independent of the energy densities. On this line of thought, the so-called cosmography approach has been proved to be an \textit{optimized} method that can reconstruct the dynamical evolution of the cosmic acceleration (via a dark energy-like term) without assuming a specific cosmological model (see \cite{Capozziello:2013wha,Bolotin:2018xtq}, and references therein). 

For several years, Cosmography has been referred to as an \textit{inverse scattering approach}, which is reasonable since it allows us to calculate cosmological constraints directly from the observational surveys by performing an expansion of the cosmological observables of interest around the present time ($a_{(z=0)}=a_0$). Nevertheless, Cosmography is not a proper inverse scattering approach since the method only circumvents the problem to fit the equation of state (EoS) parameters (with a model-dependency) by replacing it with their fit of its cosmographic parameters (with a polynomial series-dependence). Therefore, the question that we want to solve is: can we construct a \textit{new cosmography approach} where the cosmological parameters are
fitted as usual and then employ them to analyse the \textit{kinematics} via its cosmographic parameters? Moreover, can this be done
without experimenting with the traditional problem of truncation of the series. 

An interesting work that tackles a similar question was addressed in \cite{Escamilla-Rivera:2019aol}, where a straightforward cosmography approach over the EoS was improved to test cosmological  parametric versions of the dynamics of dark energy. In this approach, a homogeneous supernovae observational sample was used to perform the statistical analysis of specific dark energy parameterisations to obtain \textit{directly} the kinematics of the cosmographic parameters. The conclusion of this proposal lies in the fact that is possible to set constraints over the cosmographic parameter values using the cosmostatistics analyses of the dynamics imposed in the dark energy-like term. Also, the idea of using supernovae data\footnote{In \cite{Escamilla-Rivera:2019aol} the Joint Lightcurve Analysis (JLA) sample was used, which has extra nuisance parameters, related to the color $\alpha$ and stretch $\beta$ luminosity, in comparison to the recent Pantheon sample. In this paper we focus on using the latter sample by setting $\alpha=\beta=0$.} was 
to avoid biased results in relation 
with the Planck cosmic microwave background (CMB) temperature data, since this survey is best fitted with a $\Lambda$CDM model that is in mild tension with constraints from dynamical data probes. Some cosmography proposals available in the literature perform a test with this latter sample, which is misleading to consider since this approach is unlikely to reveal any new physics beyond an improved $\Lambda$CDM with higher corrections. And these corrections are directly linked to low redshift data precision --also in the context of a Bayesian approach--.

According to the above issues and with the results in \cite{Escamilla-Rivera:2019aol}, our proposal in this paper consists of an \textit{inverse technique} that can obtain a \textit{generic} equation of state without assuming directly a cosmographic series. Therefore, we relax two inherent cosmography problems, namely
\begin{enumerate}[label=(\roman*)]
\item The study of the dark energy dynamics (without a cosmological constant $\Lambda$) by proposing standard Taylor series or suitable Pad\'e or Chebyshev polynomials evolving with a cold dark matter (CDM) fluid. 
\item The calculations of the cosmographic parameters using (i) without truncation problems over the series/polynomials proposed.
\end{enumerate}

As an extension of this approach, and in order to work with a homogeneous data sample at low redshift, in \cite{Escamilla-Rivera:2019hqt} a novel computational tool based in machine learning (ML) for supernovae, called Recurrent-Bayesian Neural Networks (RNN+BNN), was proposed. A deep learning architecture
can develop a trained homogeneous supernovae sample in where the resulting dynamics of dark energy could lead to the necessity of another cosmological model different from $\Lambda$CDM. In such a scenario, our \textit{inverse cosmography approach} can fit statistically very well since  it
is not necessary to consider higher-order corrections of the cosmography series to obtain a convergence best fit in comparison to the standard dark energy EoSs. 
Using trained data from a deep learning architecture
for a homogeneous sample as supernovae show, from a Bayesian point of view, better evidence without dealing with the degeneracy of the models at the third order of the series. This aspect usually appears when a combination of surveys is considered \cite{Capozziello:2019cav}. 
\textbf{This evidence $\mathcal{E}=\int{\mathcal{L}(\theta) P(\theta) d\theta}$, where $\theta$ of the vector of free cosmographic parameters, $P(\theta)$ is the prior distribution of these parameters and $\mathcal{L}$ if the maximum likelihood of the model under consideration, has been computed using a MultiNest method for different activation function (see Table 6 in \cite{Escamilla-Rivera:2019hqt}). }

While in the near future, high-redshift spectroscopy surveys \cite{spec_surveys,Schlegel:2019eqc,EDGES} or Gravitational Waves surveys \cite{Corman:2020pyr,LIGOScientific:2020stg} will provide accurate data in redshifts of $2 < z <5$, our present deep learning result \cite{Escamilla-Rivera:2019hqt} allows to acquire more density of precise data of one species, the modulus distance, in redshifts range $0.01 < z < 4$.

To achieve our goal, this paper is organised as follows: in Sec.\textbf{\ref{sec:inverse_cosmo}} we describe the standard background cosmography in order to get the inverse cosmography. In Sec.\textbf{\ref{sec:eos_theory}} we present three standard cases for the EoS with dark energy-like terms $f(z)$ as cosmological fluids. Also, we introduce the description of the $f(z)$CDM model using Pad\'e and Chebyshev polynomial approaches. We calculate the EoS derived from the inverse cosmography and their cosmographic parameters at present time up to the fourth order in the series. In Sec.\textbf{\ref{sec:DL_SN}} we introduce the RNN+BNN architecture to train the supernovae Pantheon sample \cite{Escamilla-Rivera:2019hqt}. The recipe to perform this training is also explained. In Sec.\textbf{\ref{sec:results}} we perform statistical analyses for the models considered and we show the information criteria results that allow us to carry out model selection. In Sec.\textbf{\ref{sec:discussion}} we will discuss the main results. To make the treatment comprehensive, in the Appendix we present the exact expressions for the cosmographic parameters for each model under consideration. 


\section{Inverse cosmography}
\label{sec:inverse_cosmo}

We start by describing the standard cosmography equations. As will follow, this approach considers a Taylor-like expansion of the scale factor around $z=0$, which is the only parameter that can measure the Hubble flow according to the cosmological principle, therefore we can write
\begin{equation}
a(t)=1+\sum_{n=1}^{\infty}\dfrac{1}{n!}\dfrac{d^n a}{dt^n}\bigg | _{t=t_0}(t-t_0)^n\ ,
\label{eq:scalefactor}
\end{equation}
where the Hubble, deceleration, jerk and snap parameters give information of the late universe and are defined respectively as:

\begin{equation} \label{eq:cosmographic_exp}
H(t)\equiv \dfrac{1}{a}\dfrac{da}{dt} \ , \hspace{0.5cm}  q(t)\equiv -\dfrac{1}{aH^2}\dfrac{d^2a}{dt^2}\ , \hspace{0.5cm}  j(t) \equiv \dfrac{1}{aH^3}\dfrac{d^3a}{dt^3}, \ 
\hspace{0.5cm} s(t)\equiv\dfrac{1}{aH^4}\dfrac{d^4a}{dt^4}\ .
\end{equation}
Extra cosmographic parameters have longer definitions which involve higher order derivatives of $H(z)$ in a sequence given by: $\dot{a}=aH, \quad \ddot{a}=-qaH^2, \quad \dddot{a}=jaH^3, \quad d^4 a/dt^4=saH^4, \quad \text{and} \quad d^5 a/dt^5=laH^5$. 

The signs of each cosmographic parameter gives information about the cosmic kinetic scenario: the sign of the deceleration parameter $q$ indicates whether the universe is accelerating (negative sign) or decelerating (positive sign), the sign of $j$ determines the change of the universe dynamics and the value of $s$ is necessary to discriminate between evolving dark energy or $\Lambda$\footnote{An analogous approach of this \textit{snap} parameter is the $O_m$-diagnostic at second order \cite{Escamilla-Rivera:2015odt}.}. From here we notice that divergences in Taylor series are observed at $z>1$, since the expansion is performed around $z=0$, as a consequence of the convergence radius.
These cosmographic parameters are essential since their character over the sign immediately shows the kinematics of the universe. Moreover, all of them are not observable quantities, therefore we require to perform a data fit over a specific model rewritten in terms of these parameters using astrophysical observations. 

At this point, we can relax the degeneracy problem among cosmological models --certainly, with the Cosmological Principle as the backbone of this approach--, and in some manner, the hypothesis model(s) under testing will be the best one(s) once we perform the corresponding statistical cosmology over it(them). The consequence of using this approach is the high degeneracy and divergence on the terms related to the cosmographic parameters. The current arguments to defend this method remain in the idea of \textit{more data, better precision} and the divergence of the series can be alleviated per se. Even with these ideas, cosmography remains unsuitable if we continue to study it with the following methodology:
\begin{enumerate}
\item Propose a Taylor (polynomial) series for the Hubble parameter around the present time.
\item Write the luminosity distance as a function of the redshift.
\item Confront the \textit{cosmographic parameters} in the series with the current astrophysical samples.
\item Set the bounds on the cosmographic parameters with several kinds of surveys.
\end{enumerate}

The final task of this methodology remains in confronting the point that
the kinematics of the universe can be understood by setting limits on the standard cosmography using not only Taylor series \cite{Li:2019qic}, but also Pad\'e and Chebyshev polynomials to parameterise cosmic distances.
\textbf{The use of these polynomials eliminates the convergence problem at higher redshifts and therefore, the systematic errors due the truncated Taylor series} 
\cite{ Gruber:2013wua,Capozziello:2017nbu}. More interesting approaches have been done towards a \textit{cosmography equation of state} \cite{Aviles:2012ay}
or to link the late universe expansion with the early universe via a \textit{parametric cosmography} \cite{Benetti:2019gmo}. However, none of them have yet set a \textit{non-stop}-rule on the proposed cosmographic series with the use of combined observational samples. Over the general ideas up to now, a low order of the series expansion will generate 
values within the error propagation of the cosmographic parameters, while a high order of the propagation can be 
\textbf{possible by using a larger sample that includes data points at higher redshift.}
This transition of redshift regions can be treated by considering different data samples. In the end, the 
amount by which the
series will be \textit{well truncated} depends on the quality of the measurements, and even then, systematic characteristics \textbf{in the observational data} continue to give the wrong results. 

All these proposals express concern about finding not only a model-independent scenario but the endless loop tests of different kinds of polynomials in order to found a well-fitted description of the kinematics of the universe. In this direction, in \cite{Escamilla-Rivera:2019aol} a mathematical expression to obtain the EoS for a specific dynamical model was proposed. This expression allows us to obtain the cosmographic parameters without assuming \textit{directly} a cosmography-dependent polynomial series  over them. Along the rest of this paper, we are going to refer to this approach as  \textit{inverse cosmography}.

With the inverse cosmography approach, we can obtain directly the cosmographic parameter values by fitting a specific function that can replace the standard dark energy-like term without dealing with the mentioned problems, i.e.
we can constrain directly the cosmological parameters for the model and use them to compute the cosmographic parameters, e.g $q(z)$ and $j(z)$ (also $q_0$ and $j_0$) without considering higher corrections over them \cite{Capozziello:2017nbu} or perform a change of variables over the redshift to avoid divergences.

To derive the inverse cosmography, we start as usual: with the spatial flatness hypothesis on the Hubble function as
\begin{equation}\label{eq:friedmann}
\left(\frac{H(z)}{H_0}\right)^2 =\Omega_k (1+z)^2 + \Omega_{m}(1+z)^3 + \Omega_r (1+z)^4 + \Omega_{i} f_{i}(z),
\end{equation}
where curvature $\Omega_k$ and radiation $\Omega_r$ contributions can be neglected in a late universe scenario, and the density closure relation is that $\Omega_m +\Omega_i =1$. The term $\Omega_i$ is related to the standard description of the current universe dynamics once the form of $f(z)$ is given. From here we are going to refer to $i=\Lambda$ as standard dark energy scenarios and $i=p$ as polynomial scenarios with their Hubble flow function given by, respectively
\begin{eqnarray}
\left(\frac{H(z)}{H_0}\right)^2 &=& \Omega_{m}(1+z)^3 +\Omega_{\Lambda} f_{\Lambda}(z), \\
\left(\frac{H(z)}{H_0}\right)^2 &=& \Omega_{m}(1+z)^3 + \Omega_{p} f_{p}(z),  \label{eq:pol}
\end{eqnarray}
where $\Omega_\Lambda = 1-\Omega_m$ and $\Omega_{p}$ is a CDM-\textit{fluid} related to the polynomial case\footnote{This can be seen as the replace of the $\Lambda$-fluid term as it was stated in \cite{Benetti:2019gmo}.}. 

However, this approach requires a fiducial model, e.g. we can assume a flat quintessence model or a dynamical dark energy model. Therefore, we can write
a \textit{generic} expression for the cosmological EoS
where we do not impose any form of dark energy-like $\Omega_i$ by formally solve (\ref{eq:friedmann}) to obtain
\begin{equation}\label{eq:genericEoS}
w(z)= -1+\frac{1}{3} (1+z)\frac{f_{i}(z)^{\prime}}{f_{i}(z)}, 
\end{equation}
where the prime denotes $d/dz$. Notice that this \textit{generic EoS} depends solely on the form of $f_{i}(z)$. This is an interesting result since we obtain an EoS that only requires a functional form of $z$ and its derivative, e.g this function can be any polynomial at hand that allow the dynamics of a specific cosmological density.

Following the inverse cosmography idea, we can use the definitions (\ref{eq:cosmographic_exp}) and  (\ref{eq:genericEoS}) with the chain rule
$\dot{}=d/dt =-(1+z)H(z)d/dz$, to obtain our new set of cosmographic parameters in terms $H(z)$ and its derivatives:
\begin{eqnarray}
q(z)= -1 +\frac{1}{2}(1+z)\frac{[H(z)^2]'}{H(z)^2}, \label{eq:q} 
\end{eqnarray}
\begin{eqnarray}
j(z) =\frac{1}{2}(1+z)^2 \frac{[H(z)^2]''}{H(z)^2} -(1+z)\frac{[H(z)^2]'}{H(z)^2} +1, \label{eq:j}
\end{eqnarray}
\begin{eqnarray}
s(z) = -\frac{1}{6} (1+z)^3 \frac{[H(z)^2]'''}{H(z)^2} +\frac{1}{2}(1+z)^2 \frac{[H(z)^2]''}{H(z)^2} +(1+z)\frac{[H(z)^2]'}{H(z)^2} -1. \label{eq:s}
\end{eqnarray}
By solving and evaluate them at $z=0$ we get the usual cosmographic series 
\begin{eqnarray}\label{eq:Hcosmo}
H(z)&=& H_0 +\frac{dH}{dz}\bigg\rvert_{z=0} z +\frac{1}{2!}\frac{d^2 H}{dz^2}\bigg\rvert_{z=0} z^2 +\frac{1}{3!} \frac{d^3 H}{dz^3}\bigg\rvert_{z=0} z^3 +\ldots %
\end{eqnarray}
Similar expressions can be calculated by expressing everything in terms of the function normalised by the Hubble constant, $E(z)=H(z)/H_0$ and its derivatives. Nonetheless, the information we can obtain from this is exactly equivalent. 


\section{The models}
\label{sec:eos_theory}

In this work, we analyse three standard dark energy scenarios and two polynomial approaches to find directly their corresponding EoS using our generic equation (\ref{eq:genericEoS}). The first subsection is devoted to the standard $w$CDM model and Taylor-like parameterisations as Chevallier-Polarsky-Linder (CPL) and Redshift Squared (RS). In the second subsection the scenario presented will be a $f(z)$CDM-like approach based in Pad\'e (2,2) and Chebyschev (2,1) polynomial expressions. To make the treatment comprehensive, in the following sections we present the cosmographic parameters for each model at the present time. In the Appendix, we develop the exact expressions for the general cosmographic parameters for each model under consideration. 

\subsection{Standard equations of state}
\label{ssec:eos_standards}

As usual when we have to deal with a specific dark energy form with
\begin{equation}
 f_\Lambda(z)=\text{exp}\left[3\int^{z}_{0}\frac{1+w_\Lambda(\tilde{z})}{1+\tilde{z}}d\tilde{z}\right],
 \end{equation}
we do not have 
a theoretical consensus on how to choose the best $w_\Lambda(z)$, and only an optimal form that can be cosmologically viable, e.g. for quintessence models ($w_\Lambda=\text{constant}$), we have
$f_\Lambda(z)=(1+z)^{3(1+w_\Lambda)}$. For a cosmological constant if $w_\Lambda=-1$ then $f_\Lambda=1$. 
Constrictions over the free parameters in the models are a useful way to compare the relative performance of several kinds of surveys to reconstruct the cosmic expansion \cite{Escamilla-Rivera:2016qwv}. Nonetheless, a generic form of $w_\Lambda$ remains unknown.

By using (\ref{eq:q})-(\ref{eq:j})-(\ref{eq:s}), we can easily derive the cosmographic parameters without dealing with
higher derivatives for the solution of $H(z)$. Also, the cosmographic expressions obtained will show a relation between the kinematics and the dynamics of the universe.
For this task, we are going to consider the following models:

\subsubsection{$w$-constant flat cosmological case ($w$CDM)}

This model can be modeled by:
\begin{equation}
\left(\frac{H(z)}{H_0}\right)^{2}=\Omega_{m}(1+z)^3 +\Omega_{\Lambda}(1+z)^{3(1+w)},  \label{eq:friedmann_w}
\end{equation}
where $\Omega_{m}$ is the present matter density and $\Omega_{\Lambda}=(1-\Omega_{m})$ the dark energy density. According to this model, and using our inverse cosmographic parameters defined above, we compute the following cosmographic parameters:
\begin{eqnarray}
q(z)=\frac{1}{2} \left[\frac{3 w \Omega _m}{\left(\Omega _m-1\right) (z+1)^{3 w}-\Omega _m}+3 w+1\right] , \label{eq:qw}
\end{eqnarray}
\begin{eqnarray}
j(z)= \frac{[9 w (w+1)+2] \left(\Omega _m-1\right) (z+1)^{3 w}-2 \Omega _m}{2 \left(\Omega _m-1\right) (z+1)^{3
   w}-2 \Omega _m}, \label{eq:jw}
\end{eqnarray}
\begin{eqnarray}
s(z)=-\frac{[9 w (w+1)+2] \left(\Omega _m-1\right) (w-z) (z+1)^{3 w}+2 z \Omega _m}{2 (z+1) \left[\left(\Omega
   _m-1\right) (z+1)^{3 w}-\Omega _m\right]}, \label{eq:sw}
\end{eqnarray}
where we notice that for dust $w=0$: $q=1/2$, $j=1$ and $s=0$ holds for any redshift. If we consider the evaluation of these parameters at $z=0$ and for a particular value of $w_{\Lambda}=-1$ we recover the standard $\Lambda$CDM scenario: $2q_0 +j_0 +s_0=2(\Omega_{m}-1)$. 

\subsubsection{Chevallier-Polarski-Linder (CPL) case}

The evolution for this model \cite{Chevallier:2000qy,Linder:2007wa} can be
represented by two parameters that exhibit the present value of the EoS $w_0$ and its overall time evolution $w_a$:
\begin{eqnarray}\label{CPL}
\left(\frac{H(z)}{H_0}\right)^{2}&=& \Omega_m (1+z)^{3}+\Omega_{\Lambda}(1+z)^{3(1+w_0 +w_a)} e^{-\left(\frac{3w_a z}{1+z}\right)}. \label{eq:friedmann_cpl}
\end{eqnarray}
Following the above prescription, we can obtain the following cosmographic parameters at present time:
\begin{eqnarray}
q_{0} &=&\frac{\left(3 w_0+1\right) \left(\Omega _m-1\right)-\Omega _m}{2 \left(\Omega _m-1\right)-2 \Omega _m},  \label{eq:cpl_q}\\
   j_{0}&=&\frac{1}{2} \left[-\left(3 w_a+2\right) \left(\Omega _m-1\right)-9 w_0^2 \left(\Omega _m-1\right)-9 w_0
   \left(\Omega _m-1\right)+2 \Omega _m\right], \label{eq:cpl_j}\\
   s_{0} &=& \frac{\left[\left(-9 w_0-2\right) w_a-w_0 \left(9 w_0 \left(w_0+1\right)+2\right)\right] \left(\Omega
   _m-1\right)}{2 \left(\Omega _m-1\right)-2 \Omega _m}.
\end{eqnarray}

\subsubsection{Redshift squared (RS) case} 

This model \cite{Barboza:2008rh} can relax CPL parameterisation divergency at low redshift regions and can be well-behaved at $z\rightarrow -1$. The evolution of this model is given by
\begin{eqnarray}
\left(\frac{H(z)}{H_0}\right)^{2}&=&\Omega_m (1+z)^{3}+(1-\Omega_m) (1+z)^{3(1+w_0)}(1+z^2)^{\frac{3w_a}{2}}. \label{eq:BA}
 \end{eqnarray}
For this case the cosmographic parameters at present time are
\begin{eqnarray}
q_{0} &=&\frac{1}{2} \left[3 \left(\Omega _m-\left(w_0+1\right) \left(\Omega _m-1\right)\right)-2\right], \label{eq:q_rs} \\
j_{0}&=& \frac{1}{2} \left[2-3 \left(w_a+3 w_0 \left(w_0+1\right)\right) \left(\Omega _m-1\right)\right],\label{eq:j_rs} \\
s_{0}&=& -\frac{1}{2} \left[\left(-9 w_0-6\right) w_a-w_0 \left(9 w_0 \left(w_0+1\right)+2\right)\right]
   \left(\Omega _m-1\right).\\ 
   \end{eqnarray}


\subsection{$f(z)$CDM Pad\'e-like equations of state}
\label{ssec:pade}

The Pad\'e approximant for cosmographic analyses \cite{Aviles:2016wel} has been shown to have larger convergence radius in comparison to Taylor-like series, as the ones described above, and it is proving to be an optimal choice to extrapolate the analysis to higher redshifts as:
\begin{equation}
P_{(n,m)}(z)=\dfrac{\displaystyle{\sum_{i=0}^{n}a_i z^i}}{1+\displaystyle{\sum_{j=1}^{m}b_j z^j}}\,,
\label{eq:Pade}
\end{equation}
where the Pad\'e approximant $P_{(n,m)}$ of order $n/m$, defines the ratio between two standard Taylor-like series as $f(z)=\sum_{i=0}^\infty a_iz^i$. In the latter reference, a possible extension to
the analysis up to $z \sim 6$ for a Pad\'e approximants of order (2,2) was presented. In our inverse cosmography approach we are going to consider this order for the polynomial since our deep learning training data is up to this redshift range. Following the proposal in \cite{Benetti:2019gmo}, but taking into account the full definition for the Pad\'e approximant, we write a new background evolution as
\begin{equation}
f_{p}(z)= \frac{P_0+P_{1}z+P_{2}z^{2}}{1+Q_{1}z+Q_{2}z^{2}}, \quad \rightarrow \quad
P_{(2,2)}=\frac{H(z)}{H_{0}} = \frac{1+P_{1}z+P_{2}z^{2}}{1+Q_{1}z+Q_{2}z^{2}}, \label{eq:H_pade}
\end{equation}
where in the standard cosmography approach, each coefficient of the polynomial is a function that depends on the cosmographic parameters \cite{Capozziello:2019cav}, making the analysis a loop problem because of the model-dependent on every of them.

At this point, we can apply our inverse cosmography approach over a $f(z)$CDM-like model, where $\Lambda$ is replaced by a Pad\'e (2,2) polynomial in (\ref{eq:pol}) and then obtain its characteristic EoS via (\ref{eq:genericEoS})
\begin{equation} \label{eq:generic_pade}
w(z)_{\text{Pad\'e}}=-1+ \frac{(z+1) \left[P_1 \left(1-Q_2 z^2\right)+P_2 z \left(Q_1 z+2\right)-P_0 \left(2 Q_2
   z+Q_1\right)\right]}{3 \left[z \left(P_2 z+P_1\right)+P_0\right) \left(z \left(Q_2
   z+Q_1\right)+1\right]}, 
\end{equation}
from where if we consider $P_0 =1$ we obtain the EoS for Pad\'e (2,2). This term in the denominator will be important from a computational point of view, since will allow us to integrate without having divergencies at low redshift (see Sec.\ref{sec:approximants}). Notice how we recover $\Lambda$CDM at larger redshifts. According to this proposal and considering the Pad\'e (2,2) (\ref{eq:H_pade}) as $f_{p}(z)$ in (\ref{eq:pol}), we can compute the corresponding cosmographic parameters at present time using again (\ref{eq:q})-(\ref{eq:j})-(\ref{eq:s}):

\begin{eqnarray}
q_{0}&=& \frac{3 \Omega _m+\Omega _p \left(-P_1-Q_1\right)}{2 \left(\Omega _m+\Omega _p\right)}-1,\\
j_0&=& \frac{\Omega _p \left[P_1 \left(Q_1+1\right)+P_2+Q_1 \left(Q_1+1\right)-Q_2\right]}{\Omega _m+\Omega
   _p}+1, \\
   s_0 &=& 1-\frac{\Omega _m+\Omega _p \left[-\left(P_1+1\right) Q_1-P_2 Q_1
   +Q_1 \left[Q_1\left(-\left(P_1+Q_1\right)\right)-Q_1\right]+Q_2 \left(P_1+2
   Q_1\right)+Q_2\right]}{\Omega _m+\Omega _p} \nonumber \\ &&
   -\frac{\left(P_1+P_2\right) \Omega _p}{\Omega _m+\Omega _p}.
\end{eqnarray}
From these latter equations we see that for the case with $P_{i}=Q_{i}=0$, we recover a $\Lambda$CDM EoS and its respectively cosmographic parameters at present time $q_{0}=-1$, $j_{0}=1$, $s_{0}=0$. This is without calibrating the luminosity distance as it is usually done in \textit{standard cosmography}.


\subsection{$f(z)$CDM Chebyshev-like equations of state}
\label{ssec:chebyshev}

In \cite{Capozziello:2017nbu} was presented a method to optimize the standard technique of rational polynomials and consisted in defining the $(n,m)$ rational Chebyshev approximant with a coefficient $b_0\neq 0$ as 
\begin{equation}
R_{(n,m)}(z)=\dfrac{\displaystyle{\sum_{i=0}^n}\ a_i T_i(z)}{1+\displaystyle{\sum_{j=1}^m}\ b_j T_j(z)}\ ,
\label{eq:rational Chebyshev}
\end{equation}
was proposed. Applying a similar procedure used to obtain the Pad\'e approximants, we can compute the background evolution with a rational Chebyshev $R_{(2,1)}$ as
\begin{equation}
f_{p}(z)=  \frac{a_0 + a_1 z +2a_2 z^2 -a_2}{1+b_1 z}, \quad \rightarrow \quad
R_{(2,1)}=\frac{H(z)}{H_{0}} = \frac{a_3 +a_1 z +2a_2 z^2}{1+b_1 z},\label{eq:H_chebyshev} 
\end{equation}
where $a_{3}=a_0 -a_2$. Once again, in the standard cosmography approach, each coefficient $a_{i}$ and $b_{i}$ is a function that depends solely on the cosmographic parameters \cite{Capozziello:2017nbu}. To relax the model-dependent loop problem, we calculate, as we did in the Pad\'e case, our inverse cosmography approach as $f(z)$CDM-like model, where $\Lambda$ is replaced now by the rational Chebyshev $R_{(2,1)}$ in (\ref{eq:pol}) and then obtain its characteristic EoS via (\ref{eq:genericEoS})
\begin{equation}
w(z)_{\text{Chebyshev}}=-1+ \frac{\left[2 a_2 z \left(b_1 z+2\right)-a_3 b_1+a_1\right](z+1)}{3 \left(z \left(2 a_2
   z+a_1\right)+a_3\right) \left(b_1 z+1\right)}, \label{eq:generic_chebyshev_eos}
\end{equation}
from where $a_3 =a_1 / b_1$ to avoid any divergencies. We will perform a change of variable over $a_0$ once we integrate this EoS in Sec.\ref{sec:approximants}, in this manner we preserved the same expression for Chebyshev approximant (\ref{eq:H_chebyshev}).
We recover again $\Lambda$CDM at larger redshifts. Now, we can compute the corresponding cosmographic parameters at present time using (\ref{eq:q})-(\ref{eq:j})-(\ref{eq:s}):

\begin{eqnarray}
q_{0}&=& \frac{\left(a_1-a_3 b_1\right) \Omega _p+3 \Omega _m}{2 \left(a_3 \Omega _p+\Omega _m\right)}-1,\\
j_0&=& \frac{\left[a_1 \left(-b_1-1\right)+a_3 b_1 \left(b_1+1\right)+2 a_2+a_3\right] \Omega _p+\Omega
   _m}{a_3 \Omega _p+\Omega _m}, \\
s_0 &=& \frac{\left[-a_1 \left(b_1+1\right)+a_3 b_1 \left(b_1+1\right)+2 a_2+a_3\right] \Omega _p+\Omega
   _m}{a_3 \Omega _p+\Omega _m}.
\end{eqnarray}
At this point, in the cases with $a_1=a_2=a_3=0$ and $b_1=0$, we obtain a cosmology with a universe that is decelerating at the present time and with a preference for a cosmological constant. Noteworthy, this theoretical approach allows us to see that the cosmography is sensible to the values of the coefficient for this polynomial and present an anti-correlation between the cosmographic parameters. Notice that this result has been discussed in \cite{Benetti:2019gmo} after an exhaustive statistical analysis with data sets. We obtain the same result without considering cosmostatistics tools yet. We verify our arguments in Sec.\textbf{\ref{sec:results}} and extend them with deep learning methods.

\begin{figure}
\centering
\includegraphics[width=0.68\textwidth,origin=c,angle=0]{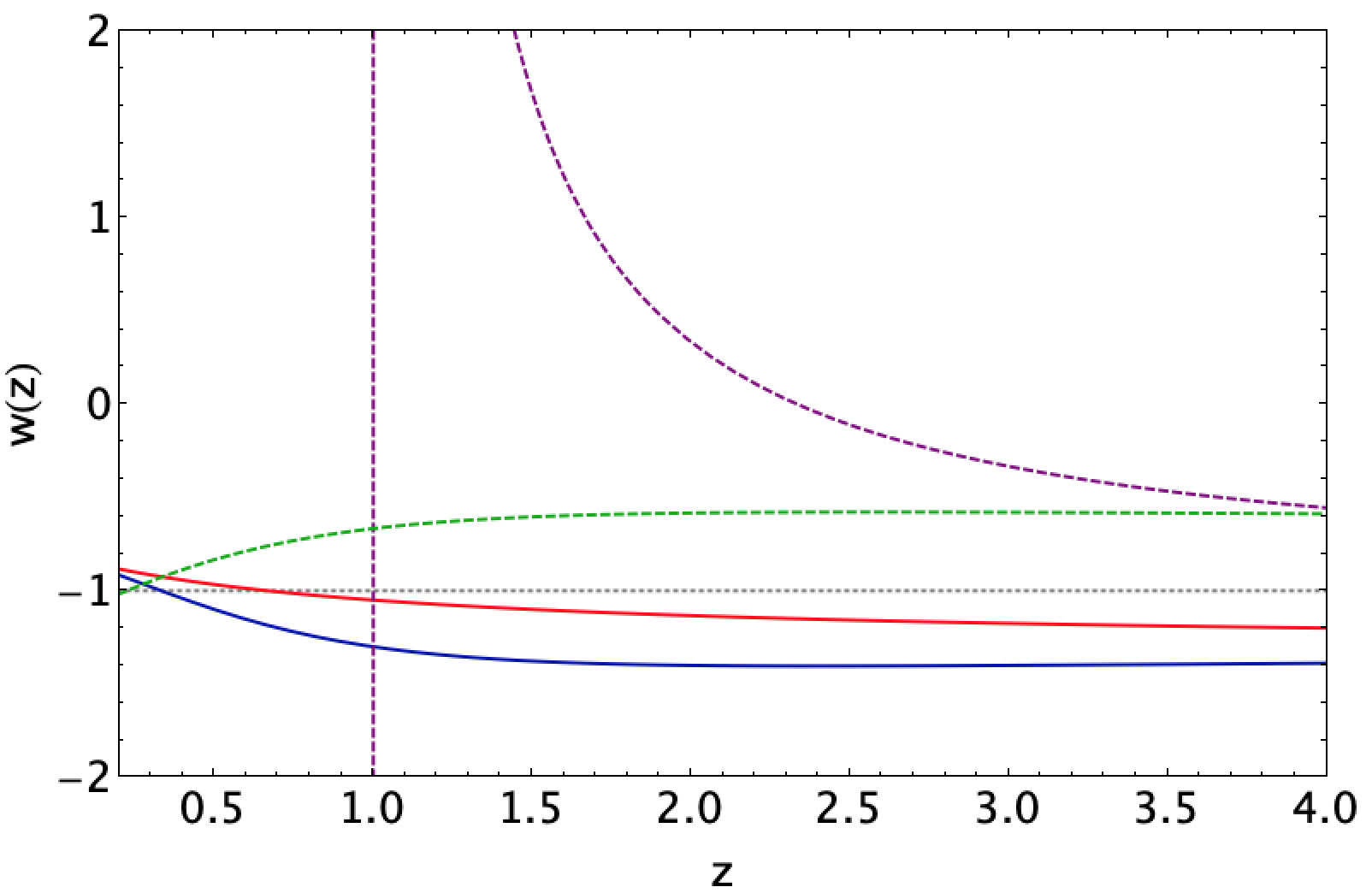}
\caption{Equation of state for the standard cosmological models Sec.(\ref{ssec:eos_standards}) and $f(z)$CDM cosmological models Secs.(\ref{ssec:pade}-\ref{ssec:chebyshev}). For this evolution, the $\Lambda$CDM (grey dotted line), the standard models CPL and RS (red and blue solid lines, respectively) and the Pad\'e (2,2) and  Chebyshev (2,1) (purple and green dashed lines, respectively) are assumed with negative values of $w_{i}$, and positive values for all the free coefficients in the polynomials.} 
\label{fig:eos_theoretical}
\end{figure}

In Fig. \ref{fig:eos_theoretical}, a comparison between our EoS is shown. According to the values considered for each of the free coefficients in the models, our analyses show a disagreement with \cite{Capozziello:2018jya} and \cite{Benetti:2019gmo}\footnote{In this reference, the authors used the same best fits values of \cite{Capozziello:2018jya}.} since the standard models behave as expected, the polynomial choices show a quintessence behaviour for $z>1$. Notice how both polynomials asymptotically approach each other near to $z\sim 4$ (our optimal redshift to perform the deep learning training, as we shall see below). Also, in the latter reference it was suggested that Pad\'e (2,2) can be a good choice in redshift regions $z<1$, but from its EoS, we obtain that at $z=1$ there is a divergence. So the \textit{link} between their toy model and the Pad\'e approach is never reached. The reason for their results lies in the use of biased best fits that experiment truncation problems to test this toy model.
 
\section{The method} 
\label{sec:DL_SN}

To understand the inverse cosmography at high redshift, we require an observational sample in a redshift range where the kinematics can offer tight constraints over the cosmological parameters of the model. The computational method adopted in this paper allows us to further extend the cosmographic fit up to $z\approx 4$, with less than 2-$\sigma$ C.L. 
We can compare this analysis with studies 
where the use of Gamma Ray Burst (GRB) data can be employed up to $z\approx6$. While exhibiting a strong correlation with their peak energy, the errors lead to biased results and the calibration at low-redshift is cosmology-dependent \cite{Vitagliano:2009et}.  

In Ref.\cite{Escamilla-Rivera:2019hqt}, one of us proposed a novel deep learning architecture to extend the observational redshift range for late-time surveys, specifically for supernovae Type Ia. The results presented offer a way to study the evolution of dark energy models from an EoS point of view. Furthermore, a modified version of this algorithm can be employed to adapt our inverse cosmography and constrain the kinematics described by the polynomials at hand. For this method, we combine two architectures: the Recurrent Neural Networks (RNN) and the Bayesian Neural Networks (BNN). From this point forward, we will refer to this method as the RNN+BNN network. In the next sections
we present our objective by describing the methodology to implement our inverse cosmography in this architecture. With this computational methodology, we can train a neural network with their confidence regions for specific homogeneous data. Additionally, RNN+BNN network minimizes the computational batch of expensive codes for dark energy models and probes the deviation from the $\Lambda$CDM scenario at large redshifts for a supernovae trained sample.
In this section, we start by describing the supernovae sample used to train the network. Afterward, we present the main methodology for each RNN and BNN neural networks to implement these cosmographic models.
Finally, we will describe the RNN+BNN architecture employed to implement the inverse cosmography. This method allows us
to constrain at high redshifts ($z=4$) the inverse cosmographic parameters, which describes the kinematical state of the Universe without dealing with the convergence (or degeneracy) problems in the polynomials.


\subsection{Supernovae sample}
The most recent supernovae Type Ia sample is considered to be trained using the combination of the two networks described in the next two subsections. The so-called
Pantheon supernovae sample 
consists of 40 bins \cite{Scolnic:2017caz} with 1048 SNIa compressed in a redshift range $[0.01, 2.3]$. Type Ia supernovae can give determinations of the distance modulus $\mu$, whose theoretical prediction is related to the luminosity distance $d_L$. As we are considering spatial flatness, the $d_L$ is related to the comoving distance $D$ through
$d_{L} (z) =c{H_0^{-1}} (1+z)D(z),$
where $c$ is the speed of light. From here we can compute
the normalised Hubble function $H(z)/H_0$ by taking the inverse of the derivative of $D(z)$ with respect to $z$
\begin{equation}
D(z)=\int^{z}_{0} \frac{H_0}{H(\tilde{z})} d\tilde{z}, \label{eq:dist}
\end{equation}
where $H_0$ is the Hubble constant prior value used to normalize the quantity $D(z)$.


\subsection{Recurrent Neural Networks (RNN) for supernovae}
To perform this first part of the neural network (NN), we consider a non-linear regression method using the supernovae sample described previously. The main idea consists in adopting a real target to train the NN for each data point. If the first data point trained is far away from the real data, then the algorithm penalizes this point and continues the process until it reaches a true value. Once the training is done for the entire sample, then the algorithm proceeds to minimise the loss function, this can be done using a Mean Squared Error (MSE) function combined with an Adam optimizer\footnote{Adam: Adaptive Moment Estimation. This method can compute learning rates for each parameter of the sample, in our case, it is restricted to the vector $(z,\mu)$. The advantage of using this method relies on the fact that it is possible to keep an exponentially decaying average of initial gradients, e.g $\mu_{t} = \beta_1 \mu_{t-1} + (1-\beta_1)g_{t}$, where $\mu(t)$ corresponds to the mean of the gradients.}.

To set the architecture for this NN, we design a cell were the output data ($\mu(z)$) of the previous step is used to calculate the new one. Each cell is provided with the information of the output value using the following equations:
\begin{eqnarray}\label{eq:info_on_in}
    h^{<t>}&=&g(W_{h}\cdot h^{<t-1>} + W_{x}\cdot x^{<t>} +b_{a}),\\
    y^{<t>}&=&g(W_{y}\cdot h^{<t>} +b_{y}),
\end{eqnarray}
where $b$ is the bias, $g$ is the activation function, $y^{<t>}$ is the output and 
$ h^{<t>}$ and $ h^{<t-1>}$, are the hidden state and its value before it, respectively. The $g$ function is given by $g = \tanh(x)$, in the range $(-1,1)$\footnote{We remark that several activations functions were used in \cite{Escamilla-Rivera:2019hqt} to train $\mu(z)$, but this Tanh offered the only physical trend according to the supernovae Ia modulus distance luminosity at recent past.}.


\subsection{Bayesian Neural Networks (BNN) for supernovae}

To solve the problem of overfitting in the networks, we perform the addition of posterior distributions making the networks probabilistic.
First, we can compute the probability distribution over a specific weight given in the distribution of the trained data and insert this as a new input for the next distribution. Finally, an entire distribution can be obtained after $n$ steps, which will increase the prediction accuracy and confidence phase spaces. The prior distribution on the weight function for a new input point $x$ can be calculated by integrating the following equation:
\begin{equation}\label{eq:prob}
    p(\mathbf{y}^{*}|\mathbf{x}^{*},\mathbf{X},\mathbf{Y})=\int p(\mathbf{y}^{*}|\mathbf{x}^{*},\mathbf{\omega})p(\mathbf{\omega}|\mathbf{X},\mathbf{Y})\mathbf{d\omega},
\end{equation}
\textbf{where \textbf{X} and \textbf{Y} denotes the redshift $z$ and the modulus distance $\mu$, respectively.}
$p(\omega| X, Y)$ is the posterior distribution over the space of parameters.
Furthermore, we can compute the model uncertainty applying this distribution in the training $n$-times. 

\begin{figure}
\centering
\includegraphics[width=1.\textwidth,origin=c,angle=0]{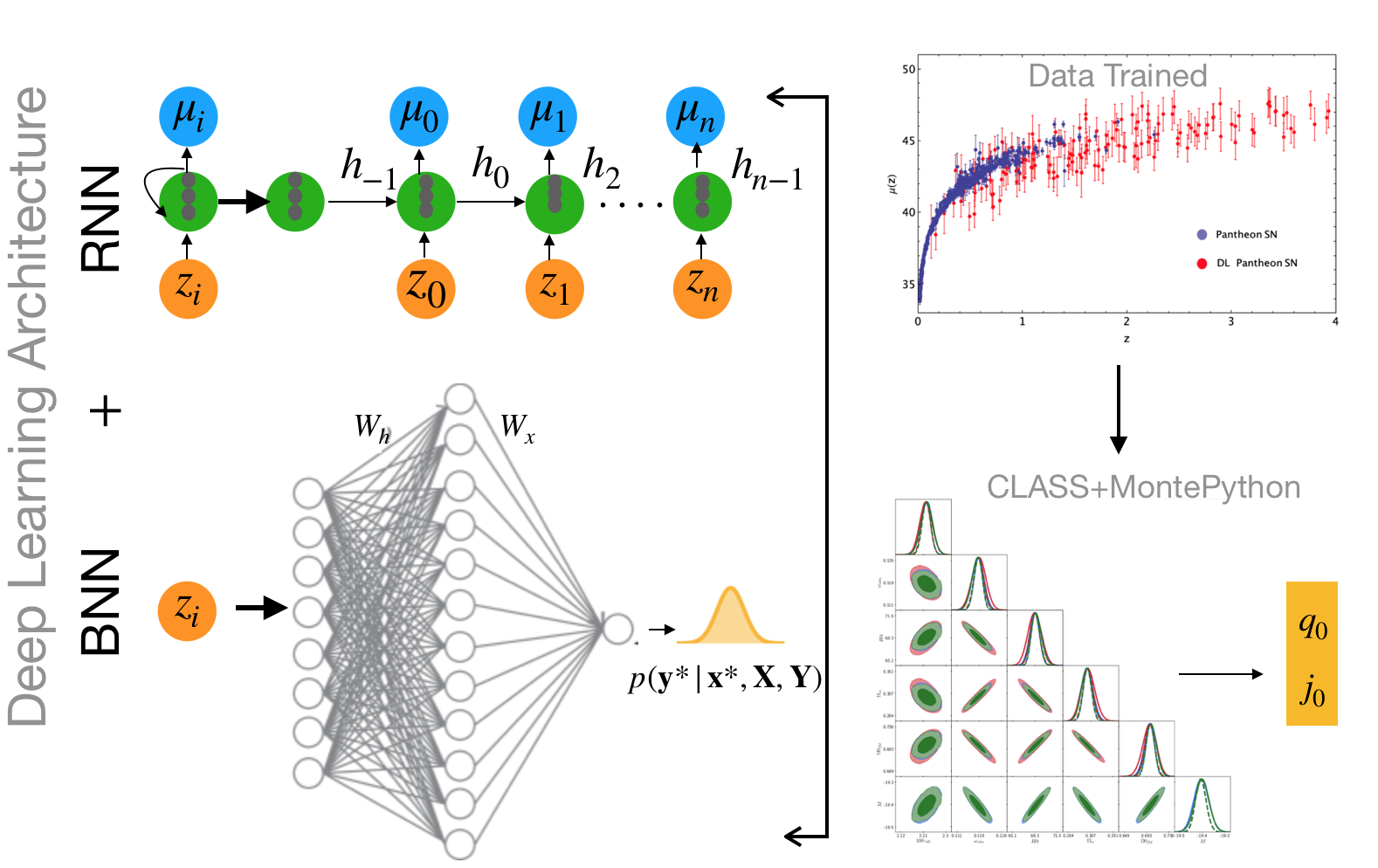}
\caption{Left Top: Deep learning RNN architecture. The input $z_i$ are the redshifts and the output $\mu_i$ is the modulus luminosity distance. In this figure we consider only one layer and the grey colored circles inside each green color node represent the number of neurons $h_i$. The values for layers and neurons are given in Step 1, Sec. \ref{sec:rnn_bnn}. Notice how each neuron receives inputs values from the latter neurons through each step $n$. In each step, an activation function Tanh is associated. Left Bottom: Deep learning BNN architecture. The network consists of one hidden layer with twelve nodes and an output layer with a single node. The weight functions between input a hidden layers are given in Eqs.(\ref{eq:info_on_in}). The result prior distribution is given by Eq.(\ref{eq:prob}).
Top Right: Supernovae final data sample trained using RNN+BNN. The RNN generates the data up to $z=4$ (see Figure \ref{fig:dl_sn}). The BNN calculates the error bars at 1-$\sigma$ C.C for the training. Bottom Right: Bayesian process using CLASS + MontePython tools to compute the cosmographic parameters best fits and their evidence using the SN final data sample.} 
\label{fig:dl_general}
\end{figure}

\subsection{RNN+BNN network for inverse cosmography}
\label{sec:rnn_bnn}

\begin{figure}
\centering
\includegraphics[width=0.5\textwidth,origin=c,angle=0]{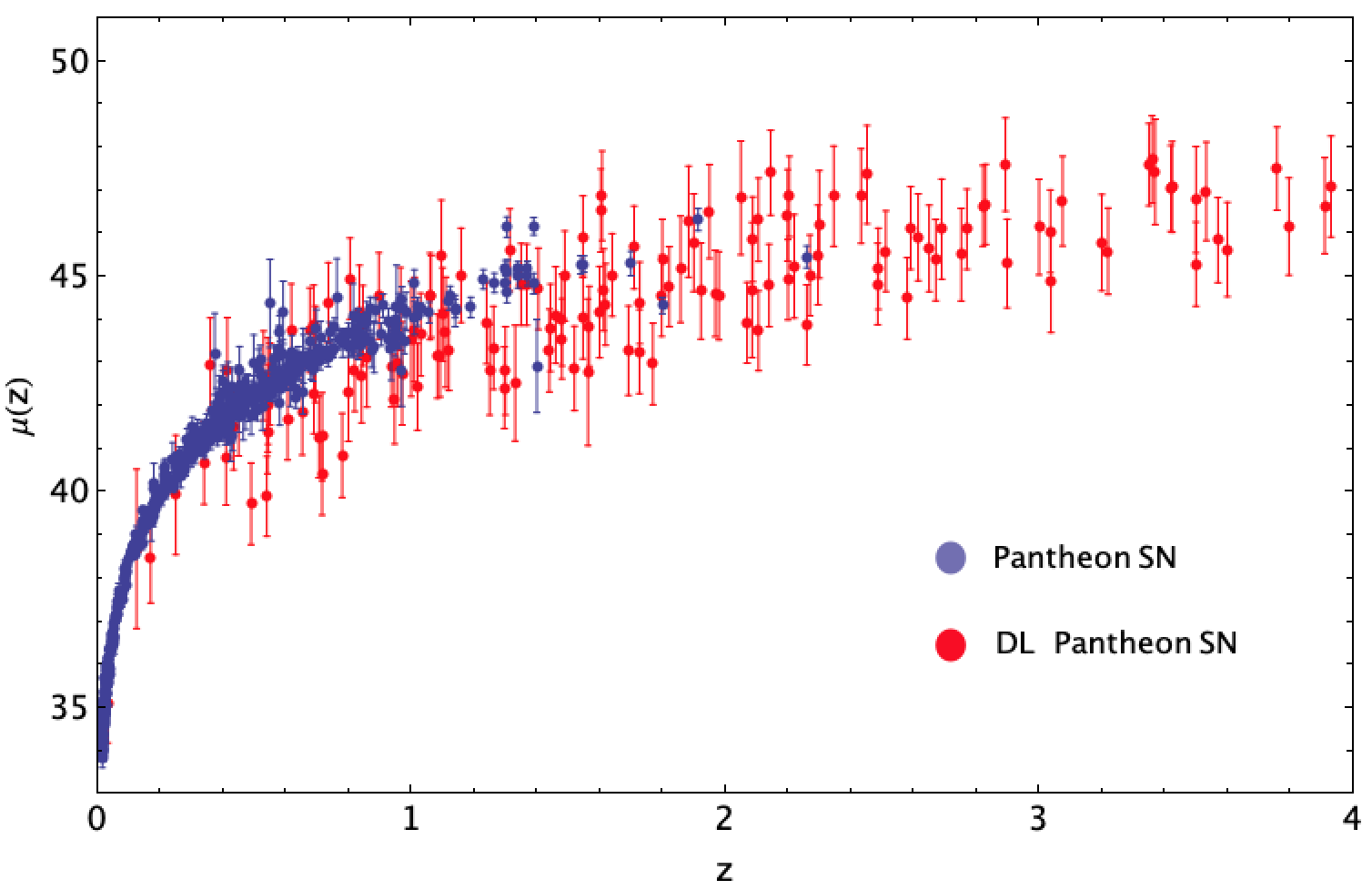}
\includegraphics[width=0.5\textwidth,origin=c,angle=0]{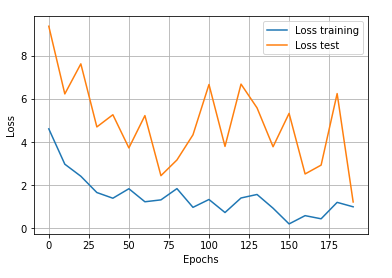}
\caption{Top: Deep learning training for Pantheon sample up to $z=4$. The error bars represent the 1-$\sigma$ C.C. The Pantheon observational data is given by the purple dots and the training RNN+BNN supernovae data is given by the red points. 
Bottom: Loss versus epochs plot for the training RNN+BNN supernovae data. This training was performed using a Tanh as activation function, which shows a better Bayesian evidence according to \cite{Escamilla-Rivera:2019hqt}.} 
\label{fig:dl_sn}
\end{figure}

To perform the deep learning training we follow the methodology given in  \cite{Escamilla-Rivera:2019hqt}. The finals steps will detail the modifications performed to insert our inverse cosmographic models. We start by dividing the process into two blocks: first, by describing the deep learning process and the characteristic on the NN architecture. Secondly, the process to compute the best fit values for the cosmographic parameters.
\begin{itemize}
    \item Training the supernovae data with RNN+BNN architecture
    \begin{itemize}
        \item Step 1. Construction of the neural network (NN). For RNN we choose a Tanh activation function with 8 hyperparameters: 
Size=4, Epochs=100, Layers=1, Neurons=100, Bathsize=10. And a variational dropout composed of an input $z$, a hidden state $h$, and an output $\mu(z)$. A diagram of this architecture was developed in \cite{Escamilla-Rivera:2019hqt}, also it demonstrated why a Tahn function is adequate to train supernovae data from a Bayesian point of view\footnote{As a summary of this process, several pieces of training with different activation functions were analised, moreover, just two of them gave a physical trend for the supernovae modulus distance $\mu(z)$, and between these, Tanh function has the better Bayesian evidence in 2-$\sigma$ C.L.  A RNN+BNN architecture portrait of these steps is given in Fig. 1 in \cite{Escamilla-Rivera:2019hqt}.}.
          \item Step 2. Organising the supernovae data. We ordering the Pantheon data from higher to low redshift with a choice of a number of steps $n=4$. 
          \textbf{We ordered in this manner since a recurrent network feed itself during the training, e.g. one network with a neuron will have a connection from the input and also from the output of its previous time step, from which feed itself. Furthermore, this will have the advantage to have more information to train in regions where the density of data points is higher (at low redshifts for this SN sample) up to regions where there is a lower density of data points (at high redshifts for this SN sample).} 
             \item Step 3. Calculation of the confidence errors via BNN. Given that the neural networks tend to overfit,
             it is important to choose a mode of regularisation. BNN allows our algorithm to calculate errors via a regularisation method. Our dropout\footnote{\textbf{Dropout is a powerful method of regularization, which randomly turns off some neurons to avoid overfitting. They can be adjusted depending on size, epochs, layers, neurons and batchsize adopted on the method.}} 
             has the following parameter: the probability to drop the input is $0$, this is because after testing several times, we found that our models could not be training with input dropout due to the loss of information. Since the cost function is MSE type (see Figure \ref{fig:dl_sn}, bottom plot) we can use Adam optimizer.
        \item Step 4. Results from the training. After the training, we read the model and apply $500$ times the same dropout to our initial model. The predicted sample with the above characteristics consist of 1810 data points in a redshift range of $[0.01-4.0]$.  See Figure \ref{fig:dl_sn}, top plot).
                \end{itemize}
     
        \item {Inverse Cosmography in Deep Learning RNN+BNN:}
        \begin{itemize}
        \item Step 5. Compute $\mu(z)$ for each cosmological model. We obtain $\mu$ by using the cosmological models described in Sec.\ref{sec:eos_theory} and integrating them to obtain $H(z)/H_0$. At this step we increase the number of epochs up to 1000.   
         \item Step 6. Calculate the best fits cosmographic parameters. We modified 
         the publicly version codes CLASS \footnote{\url{https://github.com/lesgourg/class_public}} and Monte Python \footnote{\url{https://github.com/baudren/montepython_public}} to constrain the models from Sec.\ref{sec:eos_theory} using RNN+BNN Pantheon sample from the Step 4 obtained and add them in Step 5. 
                 \end{itemize}
        \end{itemize}
        
\begin{figure}
\centering
\includegraphics[width=1.\textwidth,origin=c,angle=0]{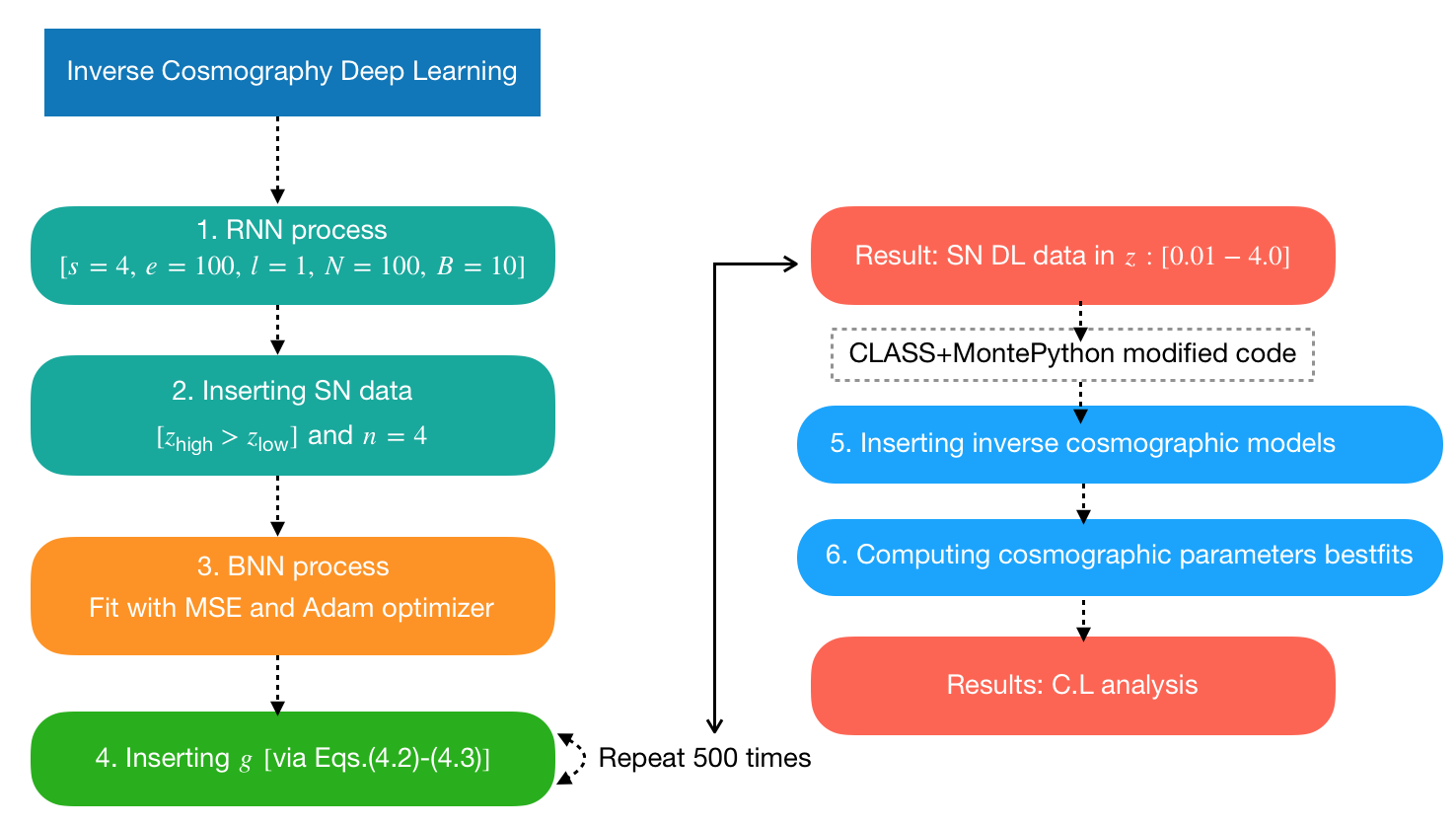}
\caption{The algorithm tree performed to obtain the cosmographic parameters best fits via inverse cosmography using the DL supernovae data trained. The full steps are described in Sec. \ref{sec:rnn_bnn}. The teal color box denotes the RNN architecture. The orange color box denotes the BNN architecture, related to the calculations of the errors. The green box denotes the inclusion of activation functions to train the observed data. Red boxes denote the final results in each tree. Blue boxes denote the process related to the standard MCMC and Bayesian calculations using CLASS+MontePython codes. Our modified version includes the models described in Eqs.(\ref{eq:friedmann_w})-(\ref{eq:friedmann_cpl})-(\ref{eq:BA})-(\ref{eq:generic_pade})-(\ref{eq:generic_chebyshev_eos}).} 
\label{fig:dl_algorithm}
\end{figure}
        
We notice that this algorithm can easily perform the fit of cosmological models with independent model priors. The only restriction 
is the
activation function, which can be viewed as a \textit{choice of a prior} since it is relevant in regions where the training data is sparse or even non-existent and the resulting trained supernovae sample depends on this choice. Nevertheless, this is a \textit{physical prior} selection since Tanh activation function can exhibit a behaviour similar to SNeIa's, where the observed differences in their peak luminosities are near correlated with observed differences in the shapes of their light curves. Also, this is supported by high Bayesian evidence  \cite{Escamilla-Rivera:2019hqt}. The numerical process is detailed in Figure \ref{fig:dl_general}. We specify the deep learning architecture, divided into the RNN (the trained data) and the BNN (the calculation of the error in the trained data). Afterward, the final sample is generated and used in the calculations of the cosmographic parameters $(q_0, j_0)$ best fits via the evidence computed by CLASS and MontePython codes. In Figure \ref{fig:dl_algorithm} we present the algorithm tree performed to obtain the cosmographic parameters best fits via inverse cosmography using the supernovae data trained. 


\section{The results: How many cosmographic parameters?}
\label{sec:results}

In this section we present the results for the information criteria AIC and BIC \cite{Liddle:2004nh} to compare our cosmological models and select the \textit{best} one in comparison to the standard $\Lambda$CDM from an inverse cosmography point of view.

In agreement with the information criteria idea, we compute the logarithm of the Bayes factor between two models $\mathcal{B}_{ij}=\mathcal{E}_{i}/\mathcal{E}_{j}$,
where the reference model ($\mathcal{E}_{i}$) with the highest evidence is the $\Lambda$CDM model and without a flat prior over $H_0$. We apply our modified version of the MCEvidence code\footnote{https://github.com/yabebalFantaye/MCEvidence} since it calculates the bayesian evidence evolving from MCMC chains employed to fit the cosmological parameters instead of fitting the cosmographic parameters as it is usually done. We based our discretization on Jeffreys's scale \cite{jeffreys}: if
$\ln{B_{ij}}<1$ there is no significant preference for  $\Lambda$CDM (or weak); if $1<\ln{B_{ij}}<2.5$ the
preference is substantial (or positive); if $2.5<\ln{B_{ij}}<5$ it is strong; if $\ln{B_{ij}}>5$ it is decisive (or very strong). \\

To develop our Bayesian analyses, we are going to consider the following steps:
\begin{enumerate}[label=(\roman*)]
\item We are using the three kinds of samples: Pantheon observational, our RNN+BNN trained supernovae sample (DL), and the join of these samples (Pantheon+DL), to perform the fitting of the cosmological parameters.
\item By statistical inference analyses, we will select the best model in comparison to $\Lambda$CDM.
\item Using the best fits values we can compute our cosmographic parameters without any degeneracy and truncation problems using our exact solutions as given in Sec.\ref{sec:eos_theory}  for each of the cosmological models. We emphasize this step since all the methodologies in the literature impose a minimum requirement to obtain a positive luminosity distance or positive $H^2$ in a redshift range $0<z/(1+z)<1$.
\end{enumerate}

\begin{table}
	\begin{center}
		
				\begin{tabular}{ |c|c|c|c| } 
	
			\hline
			Parameters & Prior CPL (uniform) & Parameters & Prior RS (uniform)\\
			\hline
			$\Omega_{m}$ & [0.25,0.35] & $\Omega_{m}$ & [0.25,0.34] \\
			\hline
			$\Omega_{\Lambda}$ & [0.65,0.75] & $\Omega_{\Lambda}$ & [0.66,0.75]  \\
			\hline
			$H_0$ & [64,74.4]  & $H_0$ & [64.8,73.7]  \\
			\hline
			M & [-19.5,-19.2] & M & [-19.5,-19.2] \\
			\hline
			$w_0$ & [-1.31,0.53]   & $w_0$ & [-1.26,-0.64]  \\
			\hline
			$w_a$ & [-0.26,0.94] & $w_a$ & [-1.36,0.46]  \\
			\hline
		\end{tabular}
	\end{center}
	\caption{Ranges of the model parameters considered in this paper. The \textit{left panel [right panel]} indicates the priors for CPL (\ref{eq:friedmann_cpl}) [RS (\ref{eq:BA})] models .}\label{tab:prior_polynomial} 
\end{table}

\begin{table}
	\begin{center}
		
				\begin{tabular}{ |c|c|c|c| } 
	
			\hline
			Parameters & Prior Pad\'e (uniform) & Parameters & Prior Chebyshev (uniform)\\
			\hline
			$\Omega_{m}$ & [0.09,0.81] & $\Omega_{m}$ & [0.09,0.89] \\
			\hline
			$\Omega_{\Lambda}$ & [0.19,0.91] & $\Omega_{\Lambda}$ & [0.10,0.90]  \\
			\hline
			M & [-19.5,-19.3] & M & [-19.5,-19.3] \\
			\hline
			$P_1$ & [0.0,1.0]   & $a_1$ & [-0.15,107]  \\
			\hline
			$P_2$ & [-0.2,0.22] & $a_2$ & [-0.72,0.61]  \\
			\hline
			$Q_1$ & [-0.3,3.0]  &$b_1$& [-12.2,18.6]\\
			\hline
			$Q_2$ & [-0.2,0.22]  &$a_3$& [-0.01,5.75]\\
			\hline
		\end{tabular}
	\end{center}
	\caption{Ranges of the model parameters considered in this paper. The \textit{left panel [right panel]} indicates the priors for Pad\'e (\ref{eq:H_pade}) [Chebyshev (\ref{eq:H_chebyshev})] models .}\label{tab:prior_polynomial2} 
\end{table}

\subsection{Standard equations of state}

\begin{itemize}
\item Case $\Lambda$CDM. We present our statistical results using Eqs.(\ref{eq:friedmann_w})-(\ref{eq:qw})-(\ref{eq:jw}) and the steps described in (i)-(ii)-(iii) above. For this model we notice that our trained data overlaps the observational results and show better confidence contour at 1-$\sigma$ (see Figure \ref{fig:bayesian_lcdm}). Also, we recover the standard cosmographic parameters values at larger redshift (up to $z=4$) as it is shown in Figure \ref{fig:cosmography_lcdm} for the three samples.

\begin{figure}
    \centering
    \includegraphics[width=0.57\textwidth,origin=c,angle=0]{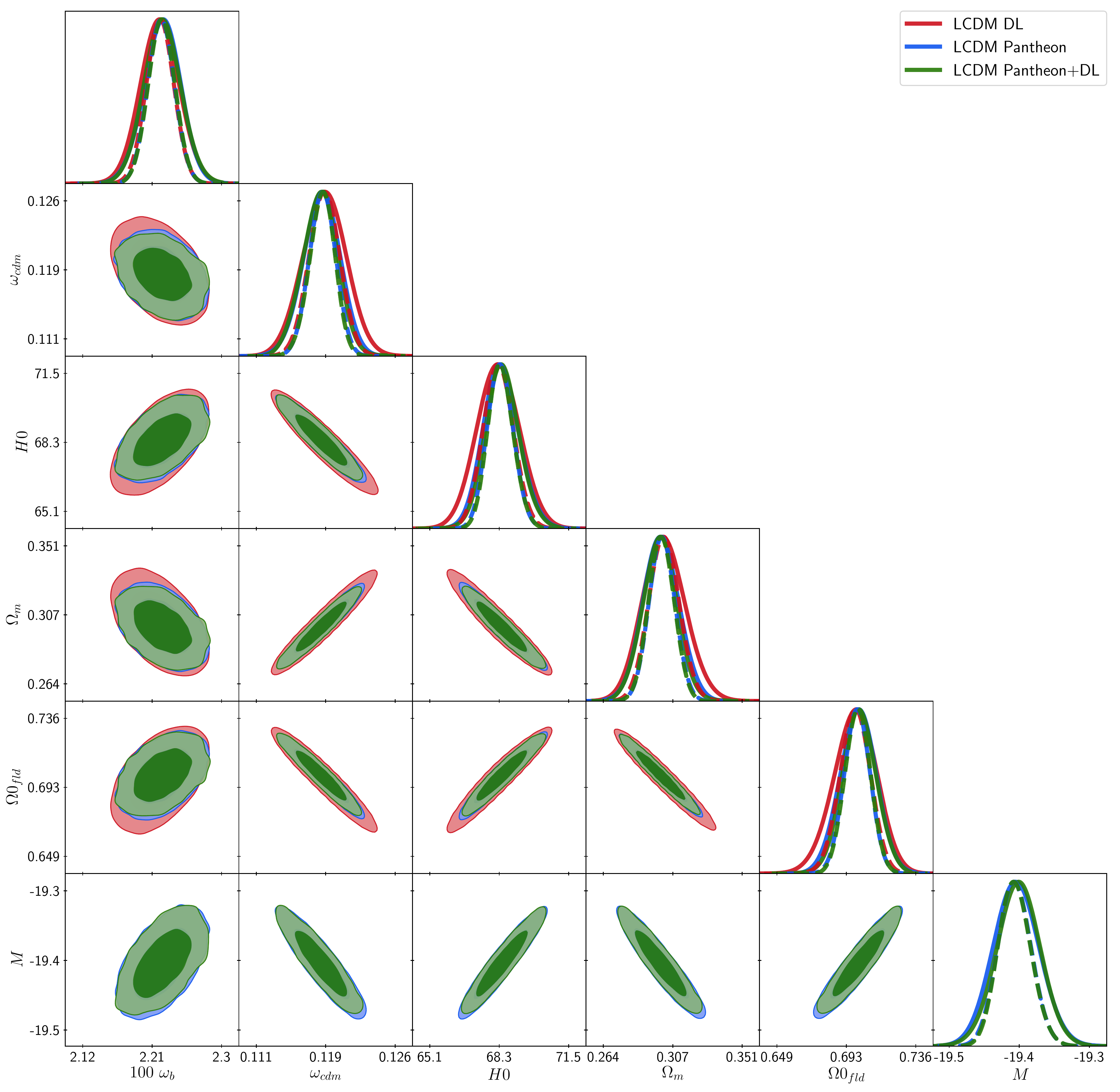}
    \caption{Confidence contours for the $\Lambda$CDM model using  Pantheon observational (blue), our RNN+BNN trained supernovae sample (DL) (red) and the join of these samples (Pantheon+DL) (green). One and two-dimensional posterior distribution are denoted by the solid line and the dashed line stands for the average likelihood distribution.}
    \label{fig:bayesian_lcdm}
\end{figure}

{\renewcommand{\tabcolsep}{6.mm}
{\renewcommand{\arraystretch}{0.5}
\begin{table}
\caption{Best fits values for $\Lambda$CDM model using Pantheon sample.}
\begin{tabular}{|l|c|c|c|c|} 
 \hline 
Parameters& Best-fit & mean$\pm\sigma$ & 95\% lower & 95\% upper \\ \hline 
$100~w_{b }$ &$2.224$ & $2.226_{-0.025}^{+0.024}$ & $2.177$ & $2.274$ \\ 
$w_{cdm }$ &$0.118$ & $0.118_{-0.002}^{+0.002}$ & $0.114$ & $0.122$ \\ 
$M$ &$-19.4$ & $-19.4_{-0.023}^{+0.022}$ & $-19.44$ & $-19.35$ \\ 
$H_0$ &$68.36$ & $68.43_{-0.88}^{+0.81}$ & $66.8$ & $70.11$ \\ 
$\Omega_{m }$ &$0.301$ & $0.300_{-0.011}^{+0.011}$ & $0.2776$ & $0.322$ \\ 
$\Omega_{\Lambda}$ &$0.700$ & $0.670_{-0.011}^{+0.011}$ & $0.678$ & $0.722$ \\ 
\hline 
 \end{tabular} \\ 
 \label{tab:lcdm_observational} \\ 
\end{table}}}

{\renewcommand{\tabcolsep}{6.mm}
{\renewcommand{\arraystretch}{0.5}
\begin{table}
\caption{Best fits values for $\Lambda$CDM model using RNN+BNN supernovae (DL) sample.}
\begin{tabular}{|l|c|c|c|c|} 
 \hline 
Parameter & Best-fit & mean$\pm\sigma$ & 95\% lower & 95\% upper \\ \hline 
$100~w_{b }$ &$2.223$ & $2.223_{-0.025}^{+0.026}$ & $2.172$ & $2.274$ \\ 
$w_{cdm }$ &$0.119$ & $0.119_{-0.0024}^{+0.0024}$ & $0.114$ & $0.123$ \\ 
$H_0$ &$68.28$ & $68.29_{-1}^{+0.97}$ & $66.33$ & $70.27$ \\ 
$\Omega_{m }$ &$0.302$ & $0.302_{-0.014}^{+0.013}$ & $0.275$ & $0.329$ \\ 
$\Omega_{\Lambda}$ &$0.698$ & $0.698_{-0.013}^{+0.014}$ & $0.671$ & $0.724$ \\ 
\hline 
 \end{tabular} \\ 
 \label{tab:lcdm_DL} \\ 
\end{table}}}

{\renewcommand{\tabcolsep}{6.mm}
{\renewcommand{\arraystretch}{0.5}
\begin{table}
\caption{Best fits values for $\Lambda$CDM model using Pantheon+(RNN+BNN) supernovae (DL) sample.}
\begin{tabular}{|l|c|c|c|c|} 
 \hline 
Parameter & Best-fit & mean$\pm\sigma$ & 95\% lower & 95\% upper \\ \hline 
$100~w_{b }$ &$2.224$ & $2.226_{-0.025}^{+0.024}$ & $2.177$ & $2.275$ \\ 
$w_{cdm }$ &$0.118$ & $0.118_{-0.002}^{+0.002}$ & $0.114$ & $0.122$ \\ 
$M$ &$-19.4$ & $-19.4_{-0.022}^{+0.021}$ & $-19.44$ & $-19.35$ \\ 
$H_0$ &$68.36$ & $68.47_{-0.84}^{+0.77}$ & $66.92$ & $70.13$ \\ 
$\Omega_{m }$ &$0.301$ & $0.299_{-0.011}^{+0.011}$ & $0.2771$ & $0.320$ \\ 
$\Omega_{\Lambda}$ &$0.699$ & $0.700_{-0.011}^{+0.011}$ & $0.680$ & $0.723$ \\ 
\hline 
 \end{tabular} \\ 
 \label{tab:lcdm_DL_obs} \\ 
\end{table}}}

\begin{figure}
    \centering
    \includegraphics[width=0.49\textwidth,origin=c,angle=0]{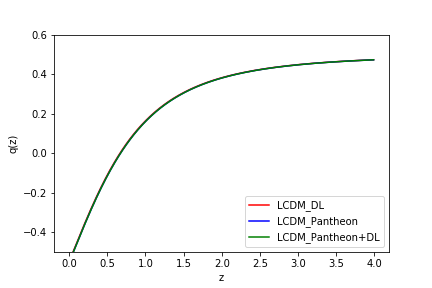}
           \includegraphics[width=0.49\textwidth,origin=c,angle=0]{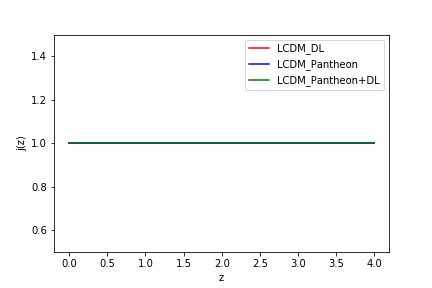}
    \caption{\textit{Left:} Cosmographic parameters for $\Lambda$CDM. \textit{Right:} Deceleration parameter $q(z)$ using Pantheon observational (blue), our RNN+BNN trained supernovae sample (DL) (red) and the join of these samples (Pantheon+DL) (green). \textit{Left:} Jerk parameter $j(z)$ using Pantheon observational (blue), our RNN+BNN trained supernovae sample (DL) (red) and the join of these samples (Pantheon+DL) (green).}
    \label{fig:cosmography_lcdm}
\end{figure}


\item Case CPL. We present our statistical results using Eqs.(\ref{eq:friedmann_cpl})-(\ref{eq:cpl_q})-(\ref{eq:cpl_j}) and the steps described in (i)-(ii)-(iii) above. For this model, we notice that our trained data overlaps the observational results in parameter spaces where there is no the dependence of the cosmological parameters. Suggesting that CPL is not a good option for $z>2.3$, the observational cutoff of Pantheon sample. When the samples are combined, the C.C is better with a correlation of the cosmological parameters at 1-$\sigma$ (see Figure \ref{fig:bayesian_cpl}). We compute the standard cosmographic parameter values at larger redshift (up to $z=4$) as it is shown in Figure \ref{fig:cosmography_cpl} for the three samples.
Notice that our RNN+BNN supernovae sample does not prefer the CPL model at low redshift, indicating that we should explore in detail another model. 
\textbf{These latter results can be seen directly from the evolution of the $w(z)$ for each data sample under consideration (see Figure \ref{fig:eos_tested}), where at low $z$, the model CPL using the RNN+BNN data trained violated the energy conditions.}
Also, this is indicated in the values of $H_0$, $\Omega_m$ and $\Omega_\Lambda$ that seems not converge (see Tables \ref{tab:CPL_observational}-\ref{tab:CPL_DL}-\ref{tab:CPL_DL_obs}). At high redshifts, both parameters recover the standard scenario.

\begin{figure}
    \centering
    \includegraphics[width=0.57\textwidth,origin=c,angle=0]{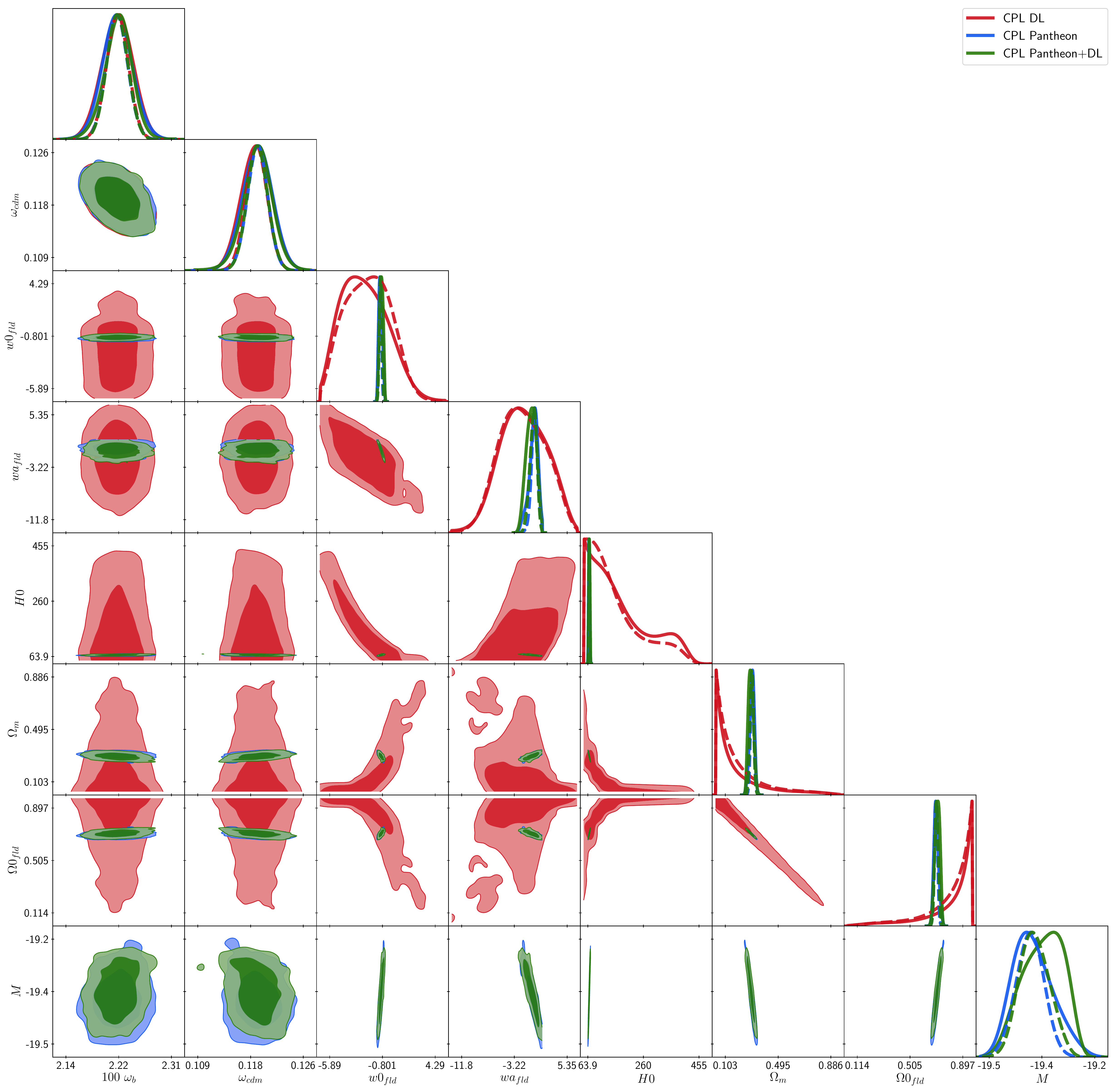}
     \caption{Confidence contours for the CPL model using  Pantheon observational (blue), our RNN+BNN trained supernovae sample (DL) (red) and the join of these samples (Pantheon+DL) (green). One and two-dimensional posterior distribution are denoted by the solid line and the dashed line stands for the average likelihood distribution.}

    \label{fig:bayesian_cpl}
\end{figure}

{\renewcommand{\tabcolsep}{6.mm}
{\renewcommand{\arraystretch}{0.5}
\begin{table}
\caption{Best fits values for CPL model using Pantheon sample.}
\begin{tabular}{|l|c|c|c|c|} 
 \hline 
Parameter & Best-fit & mean$\pm\sigma$ & 95\% lower & 95\% upper \\ \hline 
$100~w_{b }$ &$2.219$ & $2.223_{-0.026}^{+0.025}$ & $2.172$ & $2.275$ \\ 
$w_{cdm }$ &$0.119$ & $0.119_{-0.002}^{+0.002}$ & $0.114$ & $0.123$ \\ 
$w_0$ &$-1.024$ & $-0.985_{-0.17}^{+0.14}$ & $-1.284$ & $-0.641$ \\ 
$w_a$ &$0.064$ & $-0.158_{-0.54}^{+0.87}$ & $-0.698$ & $-0.711$ \\ 
$M$ &$-19.4$ & $-19.39_{-0.067}^{+0.052}$ & $-19.52$ & $-19.26$ \\ 
$H_0$ &$68.41$ & $68.64_{-2}^{+1.5}$ & $65.03$ & $72.47$ \\ 
$\Omega_{m }$ &$0.301$ & $0.230_{-0.015}^{+0.018}$ & $0.265$ & $0.334$ \\ 
$\Omega_{\Lambda}$ &$0.699$ & $0.700_{-0.018}^{+0.015}$ & $0.666$ & $0.735$ \\ 
\hline 
 \end{tabular} \\ 
 \label{tab:CPL_observational} \\ 
\end{table}}}

{\renewcommand{\tabcolsep}{6.mm}
{\renewcommand{\arraystretch}{0.5}
\begin{table}
\caption{Best fits values for CPL model using RNN+BNN supernovae (DL) sample.}
\begin{tabular}{|l|c|c|c|c|} 
 \hline 
Parameter & Best-fit & mean$\pm\sigma$ & 95\% lower & 95\% upper \\ \hline 
$100~w_{b }$ &$2.215$ & $2.223_{-0.026}^{+0.026}$ & $2.171$ & $2.274$ \\ 
$w_{cdm }$ &$0.117$ & $0.1185_{-0.0024}^{+0.002}$ & $0.1137$ & $0.123$ \\ 
$w_0$ &$0.077$ & $-2.411_{-2.1}^{+2.7}$ & $-4.511$ & $0.289$ \\ 
$w_a$ &$-0.055$ & $-1.749_{-3.8}^{+4.1}$ & $-8.729$ & $5.576$ \\ 
$H_0$ &$64.51$ & $166.2_{-1.3e+02}^{+35}$ & $37.82$ & $379.8$ \\ 
$\Omega_{m }$ &$0.335$ & $0.154_{-0.15}^{+0.002}$ & $0.004$ & $0.156$ \\ 
$\Omega_{\Lambda}$ &$0.665$ & $0.846_{-0.002}^{+0.15}$ & $0.844$ & $0.996$ \\ 
\hline 
 \end{tabular} \\ 
 \label{tab:CPL_DL} \\ 
\end{table}}}

{\renewcommand{\tabcolsep}{6.mm}
{\renewcommand{\arraystretch}{0.5}
\begin{table}
\caption{Best fits values for CPL model using Pantheon+(RNN+BNN) supernovae (DL) sample.}
\begin{tabular}{|l|c|c|c|c|}
 \hline
Parameter & Best-fit & mean$\pm\sigma$ & 95\% lower & 95\% upper \\ \hline
$100~w_{b }$ &$2.226$ & $2.223_{-0.025}^{+0.025}$ & $2.173$ & $2.273$ \\
$w_{cdm }$ &$0.118$ & $0.118_{-0.002}^{+0.003}$ & $0.113$ & $0.123$ \\
$w_0$ &$-1.006$ & $-0.9189_{-0.17}^{+0.18}$ & $-1.243$ & $-0.6056$ \\
$w_a$ &$-0.004$ & $-0.508_{-0.79}^{+0.96}$ & $-2.073$ & $0.983$ \\
$M$ &$-19.4$ & $-19.36_{-0.059}^{+0.082}$ & $-19.49$ & $-19.24$ \\
$H_0$ &$68.58$ & $69.43_{-1.9}^{+2.2}$ & $65.73$ & $73.11$ \\
$\Omega_{m }$ &$0.299$ & $0.293_{-0.02}^{+0.018}$ & $0.258$ & $0.329$ \\
$\Omega_{\Lambda }$ &$0.701$ & $0.707_{-0.018}^{+0.02}$ & $0.671$ & $0.742$ \\
\hline

 \end{tabular} \\ 
 \label{tab:CPL_DL_obs} \\ 
\end{table}}}

\begin{figure}
    \centering
    \includegraphics[width=0.49\textwidth,origin=c,angle=0]{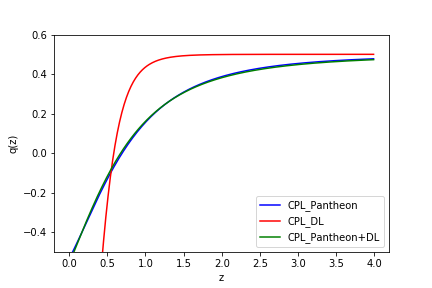}
           \includegraphics[width=0.49\textwidth,origin=c,angle=0]{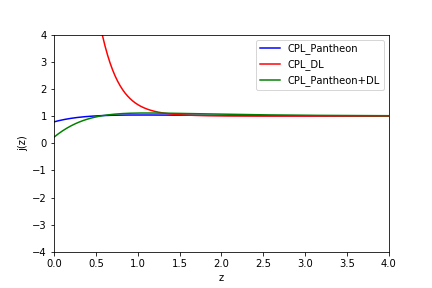}
    \caption{\textit{Left:} Cosmographic parameters for CPL. \textit{Right:} Deceleration parameter $q(z)$ using Pantheon observational (blue), our RNN+BNN trained supernovae sample (DL) (red) and the join of these samples (Pantheon+DL) (green). \textit{Left:} Jerk parameter $j(z)$ using Pantheon observational (blue), our RNN+BNN trained supernovae sample (DL) (red) and the join of these samples (Pantheon+DL) (green).}
    \label{fig:cosmography_cpl}
\end{figure}


\item Case RS. We present our statistical results using Eqs. (\ref{eq:BA})-(\ref{eq:q_rs})-(\ref{eq:j_rs}) and the steps described in (i)-(ii)-(iii) above. For this model, we notice that our trained data continue overlapping with the observational results in parameter spaces where there is no dependence on the cosmological parameters. We notice that when samples are combined, the C.C is better with a correlation of the cosmological parameters at 1-$\sigma$ (see Figure \ref{fig:bayesian_rs}).  We compute the standard cosmographic parameter values at larger redshift (up to $z=4$) as it is shown in Figure \ref{fig:cosmography_rs} for the three samples. Here we notice that our RNN+BNN supernovae sample does have a preference for RS in comparison to the CPL model at low redshift, indicating that we should explore polynomials with $z^2$-factors. This is also indicated in the values of $H_0$, $\Omega_m$ and $\Omega_\Lambda$ that seems to converge better (see Tables \ref{tab:RS_observational}-\ref{tab:RS_DL}-\ref{tab:RS_DL_obs}). At high redshifts, both parameters recover the standard scenario. Interesting enough, in both CPL and RS cases, there is a deviation transition for $j$ in $z=[0.5, 1]$, redshift regions where the Pad\'e and Chebyshev could be considered.

\begin{figure}
    \centering
    \includegraphics[width=0.57\textwidth,origin=c,angle=0]{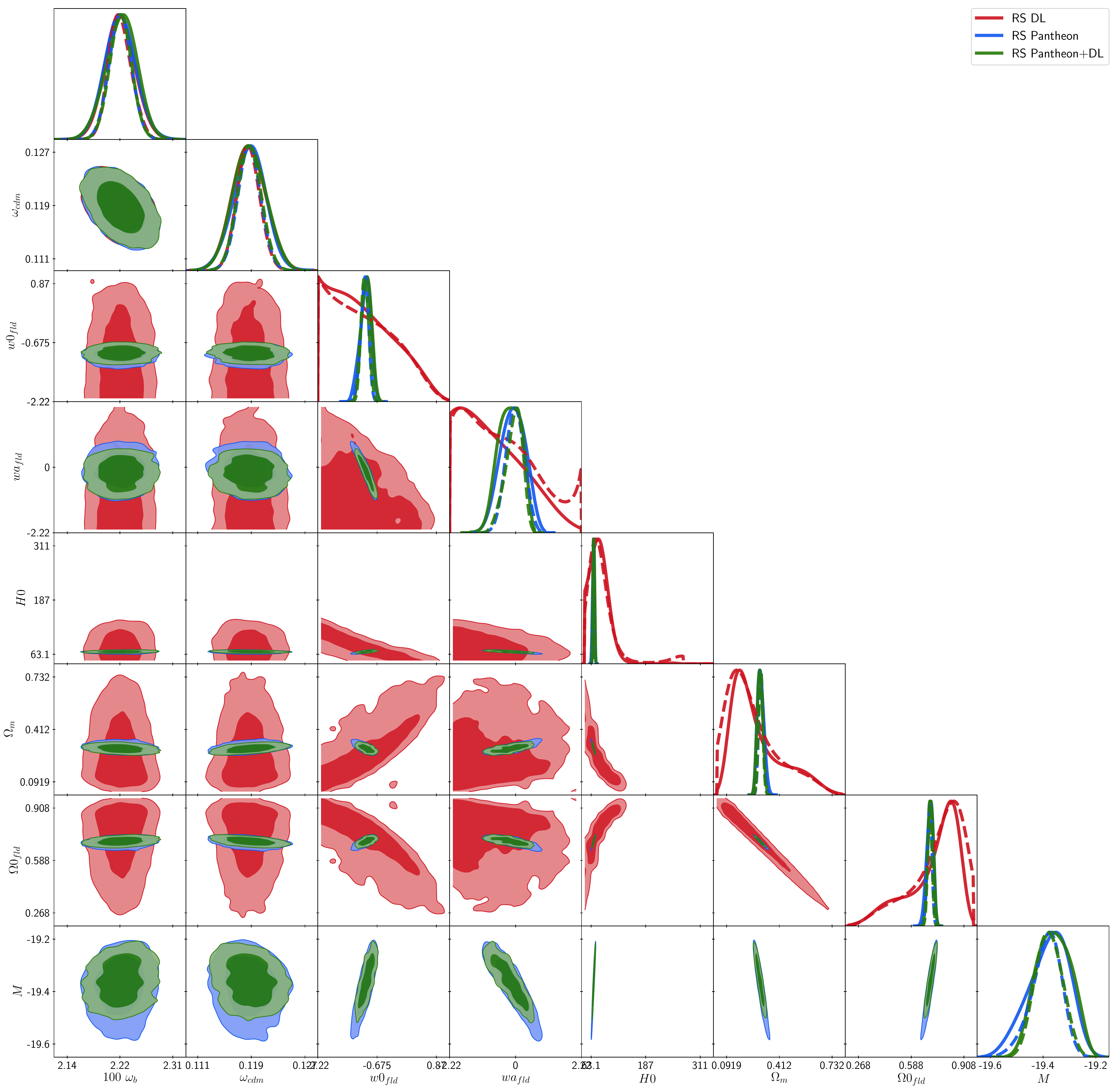}
     \caption{Confidence contours for the RS model using  Pantheon observational (blue), our RNN+BNN trained supernovae sample (DL) (red) and the join of these samples (Pantheon+DL) (green). One and two-dimensional posterior distribution are denoted by the solid line and the dashed line stands for the average likelihood distribution.}
    \label{fig:bayesian_rs}
\end{figure}

{\renewcommand{\tabcolsep}{6.mm}
{\renewcommand{\arraystretch}{0.5}
\begin{table}
\caption{Best fits values for RS model using Pantheon sample.}
\begin{tabular}{|l|c|c|c|c|} 
 \hline 
Parameter & Best-fit & mean$\pm\sigma$ & 95\% lower & 95\% upper \\ \hline 
$100~w_{b }$ &$2.224$ & $2.222_{-0.026}^{+0.026}$ & $2.171$ & $2.273$ \\ 
$w_{cdm }$ &$0.118$ & $0.119_{-0.002}^{+0.002}$ & $0.114$ & $0.123$ \\ 
$w_0$ &$-1.014$ & $-0.986_{-0.12}^{+0.16}$ & $-1.282$ & $-0.717$ \\ 
$w_a$ &$0.019$ & $-0.100_{-0.4}^{+0.42}$ & $-0.918$ & $0.726$ \\ 
$M$ &$-19.4$ & $-19.39_{-0.062}^{+0.076}$ & $-19.54$ & $-19.25$ \\ 
$H_0$ &$68.47$ & $68.67_{-2}^{+2.2}$ & $64.36$ & $72.85$ \\ 
$\Omega_{m }$ &$0.3$ & $0.299_{-0.021}^{+0.018}$ & $0.260$ & $0.340$ \\ 
$\Omega_{\Lambda}$ &$0.700$ & $0.701_{-0.018}^{+0.021}$ & $0.661$ & $0.740$ \\ 
\hline 
 \end{tabular} \\  \label{tab:RS_observational} \\ 
\end{table}}}

{\renewcommand{\tabcolsep}{6.mm}
{\renewcommand{\arraystretch}{0.5}
\begin{table}
\caption{Best fits values for RS model using RNN+BNN supernovae (DL) sample.}
\begin{tabular}{|l|c|c|c|c|} 
 \hline 
Parameter & Best-fit & mean$\pm\sigma$ & 95\% lower & 95\% upper \\ \hline 
$100~w_{b }$ &$2.22$ & $2.222_{-0.025}^{+0.025}$ & $2.172$ & $2.272$ \\ 
$w_{cdm }$ &$0.119$ & $0.119_{-0.003}^{+0.002}$ & $0.114$ & $0.123$ \\ 
$w_0$ &$-1.738$ & $-1.038_{-1.2}^{+0.36}$ & $-2.22$ & $0.4432$ \\ 
$w_a$ &$1.721$ & $-0.7272_{-1.5}^{+0.46}$ & $-2.22$ & $1.223$ \\ 
$H_0$ &$177.4$ & $81.61_{-35}^{+16}$ & $41.4$ & $126.4$ \\ 
$\Omega_{m }$ &$0.045$ & $0.274_{-0.2}^{+0.074}$ & $0.045$ & $0.611$ \\ 
$\Omega_{\Lambda}$ &$0.955$ & $0.725_{-0.074}^{+0.2}$ & $0.389$ & $0.955$ \\ 
\hline 
 \end{tabular} \\ 
 \label{tab:RS_DL} \\ 
\end{table}}}

{\renewcommand{\tabcolsep}{6.mm}
{\renewcommand{\arraystretch}{0.5}
\begin{table}
\caption{Best fits values for RS model using Pantheon+(RNN+BNN) supernovae (DL) sample.}
 \begin{tabular}{|l|c|c|c|c|}
 \hline
Parameter & Best-fit & mean$\pm\sigma$ & 95\% lower & 95\% upper \\ \hline
$100~\omega_{b }$ &$2.223$ & $2.223_{-0.026}^{+0.025}$ & $2.172$ & $2.275$ \\
$\omega_{cdm }$ &$0.119$ & $0.119_{-0.002}^{+0.003}$ & $0.114$ & $0.123$ \\
$w_0$ &$-1.014$ & $-0.950_{-0.13}^{+0.13}$ & $-1.2$ & $-0.709$ \\
$w_a$ &$0.012$ & $-0.203_{-0.31}^{+0.5}$ & $-0.923$ & $0.488$ \\
$M$ &$-19.4$ & $-19.37_{-0.040}^{+0.001}$ & $-19.41$  & $-19.36$  \\
$H_0$ &$68.47$ & $69.22_{-2.000}^{+1.900}$ & $65.87$ & $72.67$ \\
$\Omega_{m }$ &$0.301$ & $0.295_{-0.019}^{+0.017}$ & $0.262$ & $0.327$ \\
$\Omega_0$ &$0.700$ & $0.705_{-0.017}^{+0.019}$ & $0.673$ & $0.738$ \\
\hline
 \end{tabular} \\
 \label{tab:RS_DL_obs} \\ 
\end{table}}}

\begin{figure}
    \centering
    \includegraphics[width=0.49\textwidth,origin=c,angle=0]{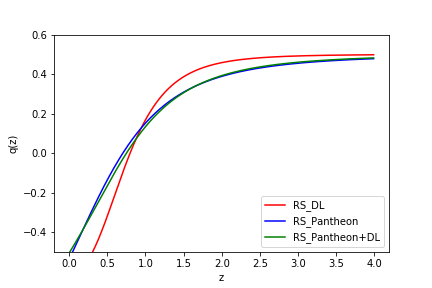}
           \includegraphics[width=0.49\textwidth,origin=c,angle=0]{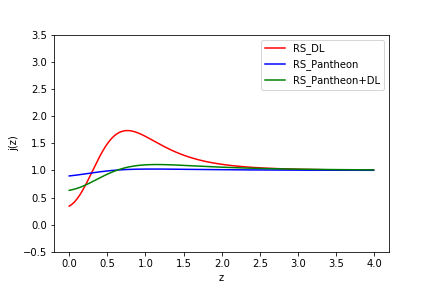}
    \caption{\textit{Left:} Cosmographic parameters for RS. \textit{Right:} Deceleration parameter $q(z)$ using Pantheon observational (blue), our RNN+BNN trained supernovae sample (DL) (red) and the join of these samples (Pantheon+DL) (green). \textit{Left:} Jerk parameter $j(z)$ using Pantheon observational (blue), our RNN+BNN trained supernovae sample (DL) (red) and the join of these samples (Pantheon+DL) (green).}
    \label{fig:cosmography_rs}
\end{figure}

\begin{figure*}
    \centering
    \includegraphics[width=0.67\textwidth,origin=c,angle=0]{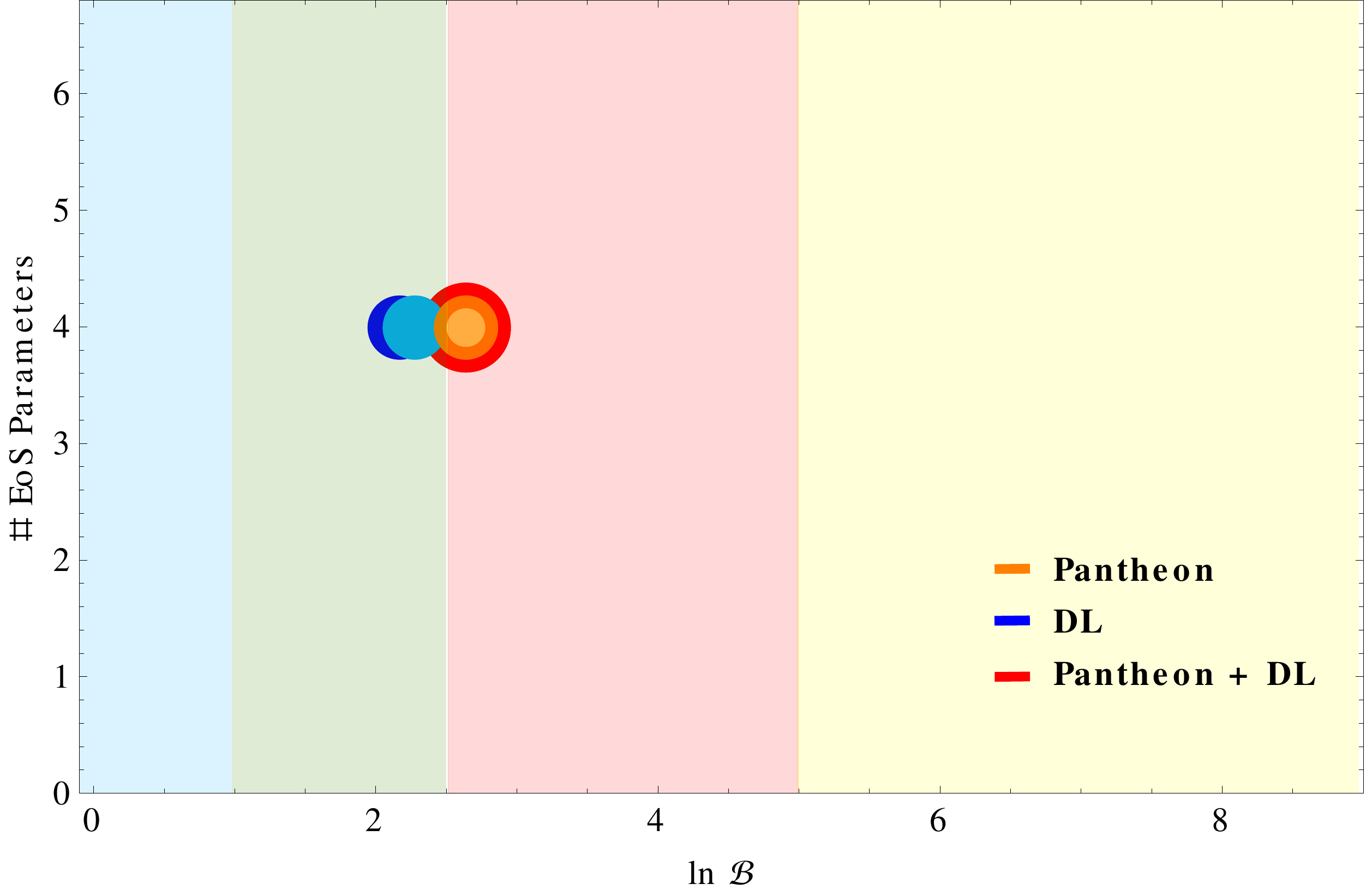}
          \caption{Bayesian portrait for CPL and RS according to the number of free EoS parameters using the values given in Tables \ref{tab:CPL_observational}-\ref{tab:CPL_DL}-\ref{tab:CPL_DL_obs} and \ref{tab:RS_observational}-\ref{tab:RS_DL}-\ref{tab:RS_DL_obs}. The qualitative color regions represent Jeffrey's scale as: strong evidence in favour of hypothesis model (light-yellow), moderate evidence in favour of hypothesis model (light-red), weak evidence for hypothesis model (light-green) and inconclusive evidence (light-blue).
          Each color in the plot legend indicates the sample used
to calculated the value $\ln B_{ij}$ in comparison to $\Lambda$CDM model. For each sample, the light color circles denote the RS model and the darker color circles denote the CPL model. This option was selected since the models overlap statistically for the observational sample and the join sample.}
    \label{fig:bayesian_comp}
\end{figure*}

\end{itemize}

\subsection{$f(z)$CDM-like equations of state}\label{sec:approximants}

To study $H_0$ tension issues using our EoS dark energy approximants-like, we consider as a fiducial prior Late Universe measurements as: $H_0= 73.8\pm 1.1$km/s/Mpc from SH0ES $+$ H0LiCOW \cite{Verde:2019ivm} in the following models:

\begin{itemize}
\item Case $f(z)$CDM Pad\'e-like. For this model, consider (\ref{eq:generic_pade}), where, as we mentioned above, $P_0 =1$ to recover Pad\'e (2,2). To perform the numerical integration of this EoS, we consider a change of variable in terms of Pad\'e approximants from $z=0$, where $P_0 = -P_1/Q_1$, and then  integrate up to $P_0=1$. In this manner we avoid any possible divergency due the Pad\'e approximant. We notice a high correlation between the approximants of the model and the cosmological parameters for our three samples, see Figure \ref{fig:bayesian_pade} and Tables \ref{tab:pade_observational}-\ref{tab:pade_DL}-\ref{tab:pade_DL_obs}. The analyses with Pantheon and DL show positive bestfits values for the approximants $P_1$ and $P_2$, which is in agreement with the theoretical analysis from Figure \ref{fig:eos_theoretical}. For the approximants $Q_1$ and $Q_2$ with contrary values in each sample cases, this is a remnant of the transition from a deceleration phase to an acceleration one as we can see from Figure \ref{fig:cosmography_pade}. We observe also that the cosmographic parameters $q$ and $j$ are indistinguishable between each other for Pantheon+DL and Pantheon samples.

\begin{figure}
    \centering
    \includegraphics[width=0.57\textwidth,origin=c,angle=0]{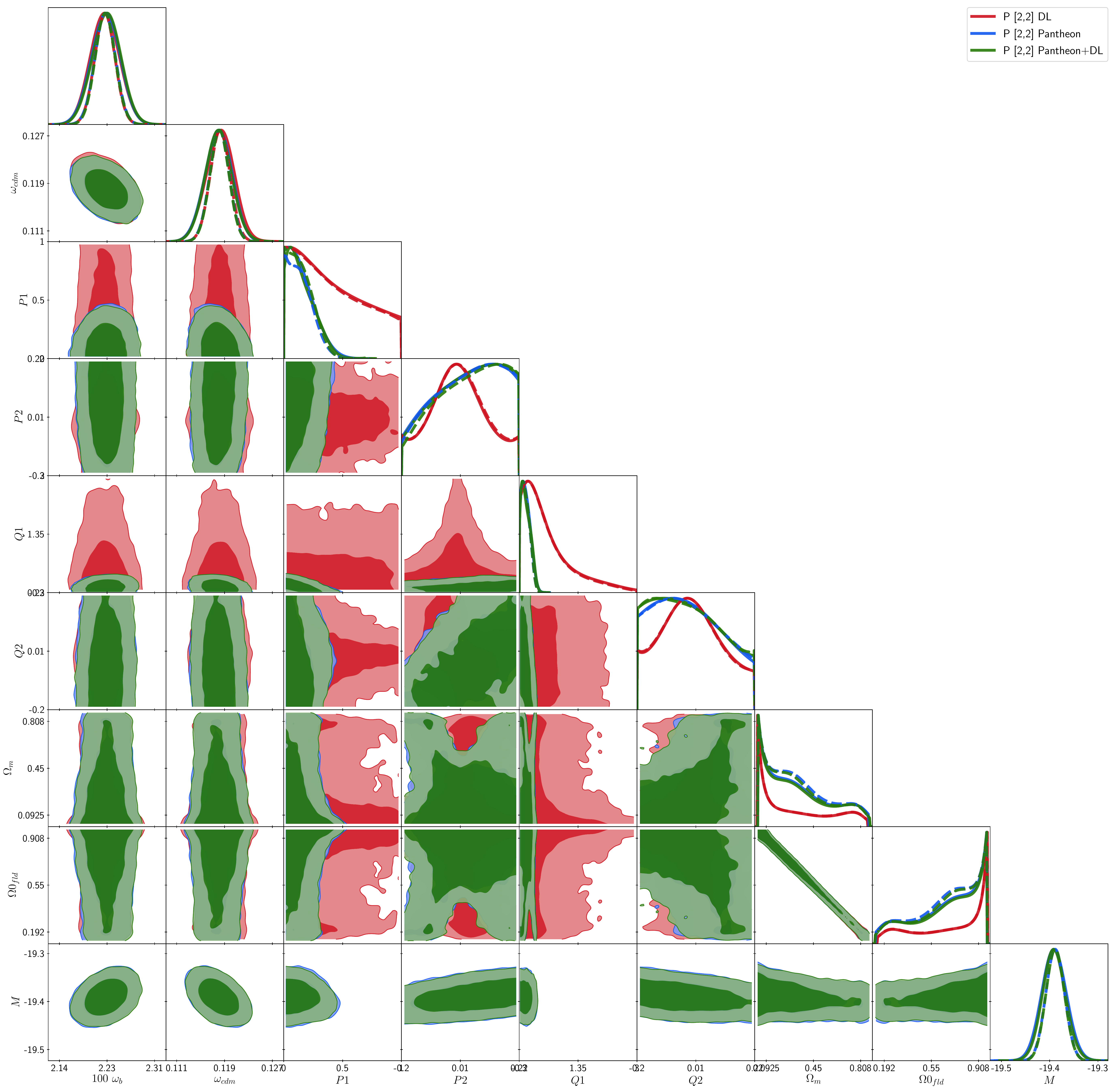}
  \caption{Confidence contours for the $f(z)$CDM Pad\'e-like model using  Pantheon observational (blue), our RNN+BNN trained supernovae sample (DL) (red) and the join of these samples (Pantheon+DL) (green). One and two-dimensional posterior distribution are denoted by the solid line and the dashed line stands for the average likelihood distribution.}
    \label{fig:bayesian_pade}
\end{figure}

{\renewcommand{\tabcolsep}{6.mm}
{\renewcommand{\arraystretch}{0.5}
\begin{table}
\caption{Best fits values for $f(z)$CDM Pad\'e-like model using Pantheon sample.}
\begin{tabular}{|l|c|c|c|c|} 
 \hline 
Parameters& Best-fit & mean$\pm\sigma$ & 95\% lower & 95\% upper \\ \hline 
$100~w_{b }$ &$2.222$ & $2.225_{-0.025}^{+0.026}$ & $2.174$ & $2.275$ \\ 
$w_{cdm }$ &$0.118$ & $0.118_{-0.002}^{+0.002}$ & $0.114$ & $0.123$ \\ 
$P_1$ &$0.183$ & $0.156_{-0.16}^{+0.043}$ & $1.462e-06$ & $0.376$ \\ 
$P_2$ &$0.040$ & $0.035_{-0.061}^{+0.18}$ & $-0.162$ & $0.22$ \\ 
$Q_1$ &$-0.204$ & $-0.107_{-0.19}^{+0.053}$ & $-0.3$ & $0.131$ \\ 
$Q_2$ &$0.0262$ & $-0.013_{-0.19}^{+0.06}$ & $-0.2$ & $0.186$ \\ 
$M$ &$-19.4$ & $-19.4_{-0.029}^{+0.028}$ & $-19.46$ & $-19.34$ \\ 
$\Omega_{m }$ &$0.251$ & $0.311_{-0.31}^{+0.093}$ & $0.001$ & $0.404$ \\ 
$\Omega_0$ &$0.749$ & $0.688_{-0.093}^{+0.31}$ & $0.595$ & $0.998$ \\ 
\hline 
 \end{tabular} \\ 
 \label{tab:pade_observational} \\ 
\end{table}}}

{\renewcommand{\tabcolsep}{6.mm}
{\renewcommand{\arraystretch}{0.5}
\begin{table}
\caption{Best fits values for $f(z)$CDM Pad\'e-like using RNN+BNN supernovae (DL) sample.}
\begin{tabular}{|l|c|c|c|c|} 
 \hline 
Parameter & Best-fit & mean$\pm\sigma$ & 95\% lower & 95\% upper \\ \hline 
$100~w_{b }$ &$2.223$ & $2.224_{-0.025}^{+0.025}$ & $2.173$ & $2.274$ \\ 
$w_{cdm }$ &$0.118$ & $0.118_{-0.002}^{+0.002}$ & $0.114$ & $0.123$ \\ 
$P_1$ &$0.089$ & $0.411_{-0.41}^{+0.14}$ & $7.315e-06$ & $0.921$ \\ 
$P_2$ &$0.151$ & $0.004_{-0.11}^{+0.099}$ & $-0.188$ & $0.196$ \\ 
$Q_1$ &$-0.222$ & $0.506_{-0.81}^{+0.14}$ & $-0.3$ & $2.07$ \\ 
$Q_2$ &$-0.137$ & $-0.004_{-0.14}^{+0.094}$ & $-0.2$ & $0.183$ \\ 
$\Omega_{m }$ &$0.0142$ & $0.28_{0.001}^{+0.002}$ & $0.281$ & $0.282$ \\ 
$\Omega_0 $ &$0.986$ & $0.720_{0.001}^{+0.002}$ & $0.721$ & $0.722$ \\ 
\hline 
 \end{tabular} \\ 
 \label{tab:pade_DL} \\ 
\end{table}}}

{\renewcommand{\tabcolsep}{6.mm}
{\renewcommand{\arraystretch}{0.5}
\begin{table}
\caption{Best fits values for $f(z)$CDM Pad\'e-like model using Pantheon+(RNN+BNN) supernovae (DL) sample.}
\begin{tabular}{|l|c|c|c|c|} 
 \hline 
Parameter & Best-fit & mean$\pm\sigma$ & 95\% lower & 95\% upper \\ \hline 
$100~w_{b }$ &$2.221$ & $2.226_{-0.026}^{+0.025}$ & $2.176$ & $2.276$ \\ 
$w_{cdm }$ &$0.119$ & $0.118_{-0.002}^{+0.002}$ & $0.114$ & $0.123$ \\ 
$P_1$ &$0.108$ & $0.154_{-0.15}^{+0.043}$ & $1.897e-05$ & $0.366$ \\ 
$P_2$ &$-0.007$ & $0.0370_{-0.06}^{+0.18}$ & $-0.161$ & $0.22$ \\ 
$Q_1$ &$-0.163$ & $-0.104_{-0.2}^{+0.055}$ & $-0.3$ & $0.135$ \\ 
$Q_2$ &$0.055$ & $-0.013_{-0.19}^{+0.061}$ & $-0.203$ & $0.048$ \\ 
$M$ &$-19.4$ & $-19.4_{-0.028}^{+0.027}$ & $-19.45$ & $-19.34$ \\ 
$\Omega_{m }$ &$0.424$ & $0.310_{-0.3}^{+0.092}$ & $0.01$ & $0.402$ \\ 
$\Omega_0$ &$0.575$ & $0.690_{-0.092}^{+0.3}$ & $0.600$ & $0.990$ \\ 
\hline 
 \end{tabular} \\ 
 \label{tab:pade_DL_obs} \\ 
\end{table}}}

\begin{figure}
    \centering
    \includegraphics[width=0.49\textwidth,origin=c,angle=0]{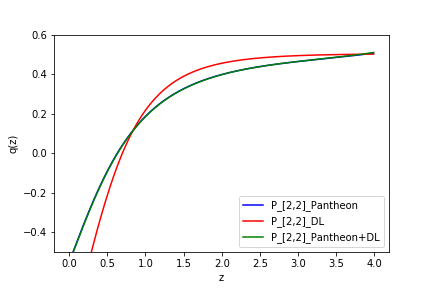}
           \includegraphics[width=0.49\textwidth,origin=c,angle=0]{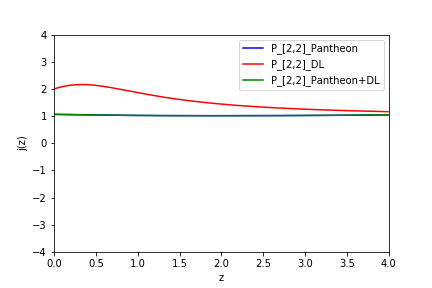}
    \caption{\textit{Left:} Cosmographic parameters for $f(z)$CDM Pad\'e. \textit{Right:} Deceleration parameter $q(z)$ using Pantheon observational (blue), our RNN+BNN trained supernovae sample (DL) (red) and the join of these samples (Pantheon+DL) (green). \textit{Left:} Jerk parameter $j(z)$ using Pantheon observational (blue), our RNN+BNN trained supernovae sample (DL) (red) and the join of these samples (Pantheon+DL) (green).}
    \label{fig:cosmography_pade}
\end{figure}


\item Case $f(z)$CDM Chebyshev-like. As in the latter case, for this model, consider (\ref{eq:generic_chebyshev_eos}), where $a_3 =a_1/b_1$, to recover Chebyshev (2,1). To perform the numerical integration of the EoS, we consider a change of variable in terms of Chebyshev approximants from $a_0 =0$, up to $a_2=-a_1/b_1$. In this manner we avoid any possible divergence and recover the original Chebyshev approximant. 
The results for this case are reported in Tables \ref{tab:chebyshev_observational}-\ref{tab:chebyshev_DL}-\ref{tab:chebyshev_DL_obs} and Figure \ref{fig:bayesian_chebyshev}. For this kind of polynomial $R_{(2,1)}$ we found a high correlation between approximants, but the confidence regions for the DL simulations gets to reproduce the same observational trend at $95\%$. 
The analyses with Pantheon and DL show a couple of positive bestfits values for the approximants $a_1$ and $a_2$ and negative values for $b_1$, in agreement with the theoretical analysis from Figure \ref{fig:eos_theoretical}. From Figure \ref{fig:cosmography_chebyshev} we notice deviations from $q$ and $j$ in comparison to the observational and join sample, but at high redshift, the three tested samples for this model are asymptotically approaching. 

\begin{figure}
    \centering
    \includegraphics[width=0.57\textwidth,origin=c,angle=0]{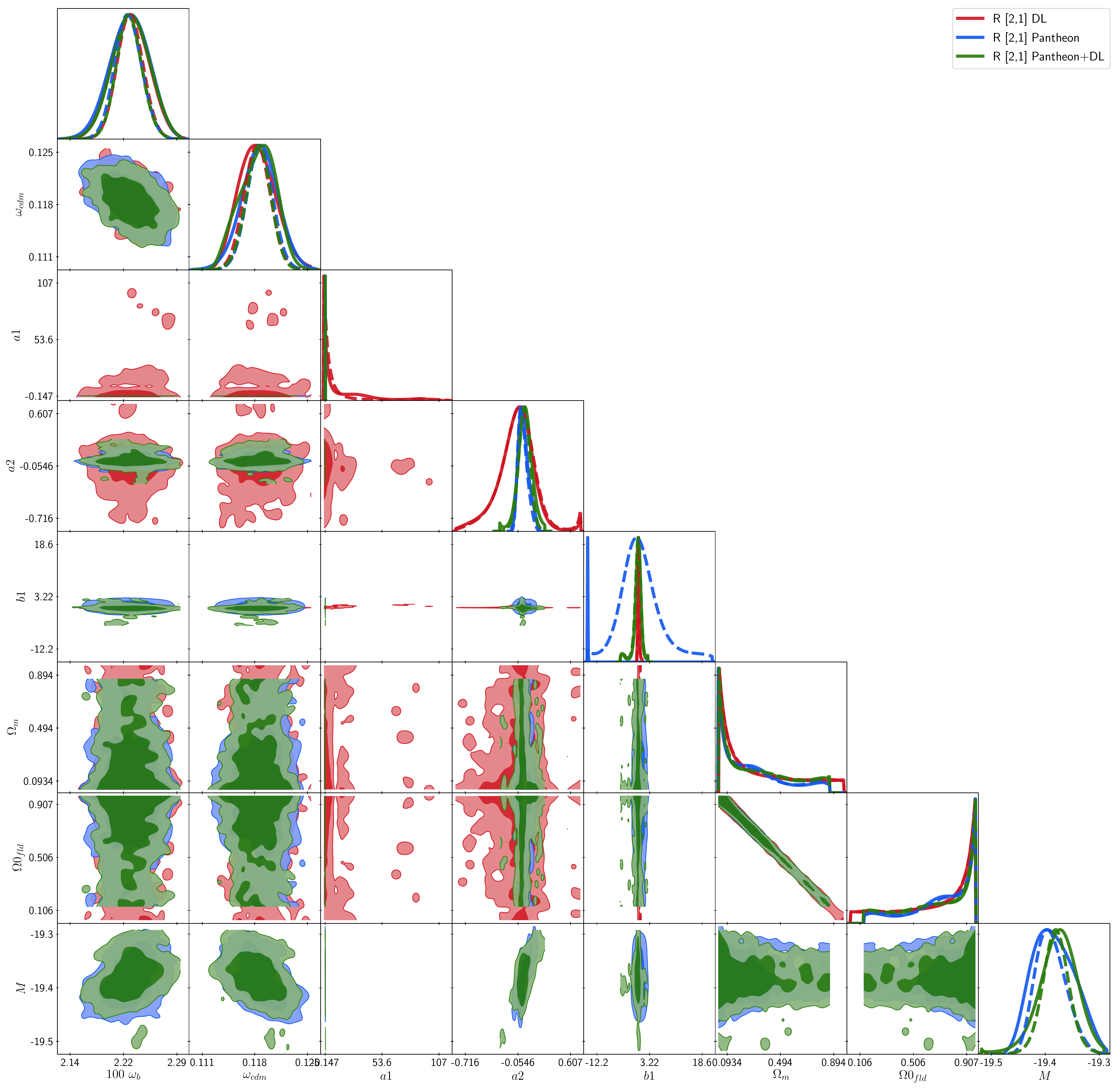}
      \caption{Confidence contours for the $f(z)$CDM Chebyshev-like model using  Pantheon observational (blue), our RNN+BNN trained supernovae sample (DL) (red) and the join of these samples (Pantheon+DL) (green). One and two-dimensional posterior distribution are denoted by the solid line and the dashed line stands for the average likelihood distribution.}
    \label{fig:bayesian_chebyshev}
\end{figure}

{\renewcommand{\tabcolsep}{6.mm}
{\renewcommand{\arraystretch}{0.5}
\begin{table}
\caption{Best fits values for $f(z)$CDM Chebyshev-like model using Pantheon sample.}
\begin{tabular}{|l|c|c|c|c|} 
 \hline
Parameter & Best-fit & mean$\pm\sigma$ & 95\% lower & 95\% upper \\ \hline 
$100~w_{b }$ &$2.222$ & $2.223_{-0.026}^{+0.028}$ & $2.168$ & $2.277$ \\
$w_{cdm }$ &$0.119$ & $0.119_{-0.002}^{+0.002}$ & $0.114$ & $0.123$ \\
$a_1$ &$0.026$ & $0.013_{-0.065}^{+0.055}$ & $-0.102$ & $0.134$ \\
$a_2$ &$-0.009$ & $0.016_{-0.002}^{+0.001}$ & $0.014$ & $0.017$ \\
$b_1$ &$0.587$ & $0.140_{-0.73}^{+1.1}$ & $-1.632$ & $2.252$ \\
$M$ &$-19.4$ & $-19.39_{-0.005}^{+0.001}$ & $-19.39$ & $-19.39$ \\
$\Omega_{m }$ &$0.134$ & $0.248_{-0.005}^{+0.001}$ & $0.243$ & $0.245$ \\
$\Omega_0$ &$0.866$ & $0.752_{-0.005}^{+0.001}$ & $0.745$ & $0.753$ \\
\hline
 \end{tabular} \\ 
 \label{tab:chebyshev_observational} \\ 
\end{table}}}

{\renewcommand{\tabcolsep}{6.mm}
{\renewcommand{\arraystretch}{0.5}
\begin{table}
\caption{Best fits values for $f(z)$CDM Chebyshev-like using RNN+BNN supernovae (DL) sample.}
\begin{tabular}{|l|c|c|c|c|} 
 \hline
Parameter & Best-fit & mean$\pm\sigma$ & 95\% lower & 95\% upper \\ \hline 
$100~w_{b }$ &$2.22$ & $2.225_{-0.024}^{+0.027}$ & $2.175$ & $2.278$ \\
$w_{cdm }$ &$0.119$ & $0.118_{-0.003}^{+0.002}$ & $0.114$ & $0.123$ \\
$a_1$ &$-0.616$ & $3.796_{-8.2}^{+2.8}$ & $-4.404$ & $6.596$ \\
$a_2$ &$0.599$ & $-0.059_{-0.13}^{+0.21}$ & $-0.188$ & $0.152$ \\
$b_1$ &$-0.027$ & $0.038_{-0.2}^{+0.14}$ & $-0.162$ & $0.178$ \\
$\Omega_{m }$ &$0.256$ & $0.291_{-0.002}^{+0.001}$ & $0.289$ & $0.292$ \\
$\Omega_0$ &$0.744$ & $0.708_{-0.002}^{+0.001}$ & $0.706$ & $0.709$ \\
\hline
 \end{tabular} \\ 
 \label{tab:chebyshev_DL} \\ 
\end{table}}}

{\renewcommand{\tabcolsep}{6.mm}
{\renewcommand{\arraystretch}{0.5}
\begin{table}
\caption{Best fits values for $f(z)$CDM Chebyshev-like model using Pantheon+(RNN+BNN) supernovae (DL) sample.}
\begin{tabular}{|l|c|c|c|c|} 
 \hline
Parameter & Best-fit & mean$\pm\sigma$ & 95\% lower & 95\% upper \\ \hline 
$100~w_{b }$ &$2.225$ & $2.225_{-0.028}^{+0.024}$ & $2.175$ & $2.279$ \\
$w_{cdm }$ &$0.119$ & $0.118_{-0.002}^{+0.002}$ & $0.114$ & $0.123$ \\
$a_1$ &$0.017$ & $0.009_{-0.079}^{+0.07}$ & $-0.149$ & $0.162$ \\
$a_2$ &$0.001$ & $0.021_{-0.064}^{+0.046}$ & $-0.119$ & $0.251$ \\
$b_1$ &$-1.618$ & $-0.360_{-0.41}^{+0.71}$ & $-0.77$ & $0.350$ \\
$M$ &$-19.4$ & $-19.39_{-0.028}^{+0.026}$ & $-19.44$ & $-19.32$ \\
$\Omega_{m }$ &$0.262$ & $0.278_{-0.001}^{+0.001}$ & $0.277$ & $0.279$ \\
$\Omega_0$ &$0.738$ & $0.722_{-0.001}^{+0.001}$ & $0.720$ & $0.722$ \\
\hline
 \end{tabular} \\ 
 \label{tab:chebyshev_DL_obs} \\ 
\end{table}}}

\begin{figure}
    \centering
    \includegraphics[width=0.49\textwidth,origin=c,angle=0]{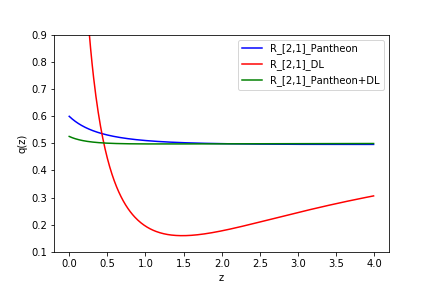}
           \includegraphics[width=0.49\textwidth,origin=c,angle=0]{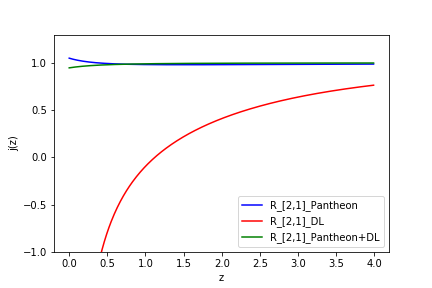}
    \caption{\textit{Left:} Cosmographic parameters for $f(z)$CDM Chebyshev. \textit{Right:} Deceleration parameter $q(z)$ using Pantheon observational (blue), our RNN+BNN trained supernovae sample (DL) (red) and the join of these samples (Pantheon+DL) (green). \textit{Left:} Jerk parameter $j(z)$ using Pantheon observational (blue), our RNN+BNN trained supernovae sample (DL) (red) and the join of these samples (Pantheon+DL) (green).}
    \label{fig:cosmography_chebyshev}
\end{figure}

\end{itemize}


\begin{table}[]
\centering
\begin{tabular}{|c |c |c |c |c |c |c|} 
\hline
\bf{Model}& $\chi^2$& $\bf{q_0}$ & $\bf{j_0}$ & \bf{$\ln{B_{ij}}$}   & \bf{AIC}   & \bf{BIC}  \\
\hline\hline
$\Lambda$CDM      &     1028      &       $ -0.551 \pm 0.014$                                    &       $1.0 $   &          -  &    -        &    -           \\
\hline
CPL           &        1028       &      $-0.534 \pm 0.308$                                  &     $0.790  \pm   1.011$    &  2.637          &         1036           &    1055.4       \\
\hline
RS            &        1028       &       $-0.536 \pm 0.335$                                    &       $0.9 \pm 1.312$    &      2.637      &           1036            &     1055.4   \\
\hline
$f(z)$CDM Pad\'e-like      &    1028         &     $-0.550 \pm 0.354$                 &   $1.063 \pm 0.282$         & 2.637           &  1036        & 1055.94 \\    
\hline
$f(z)$CDM Chebyshev-like      &     1028        &     $0.599\pm 0.184$              & $1.050\pm 0.460$           &  3.042          &    1038      & 1062.92 \\                  
\hline
\end{tabular}
\caption{Cosmological models and approximants using Pantheon sample. Column 1: cosmographic parameters values at 2-$\sigma$. Column 2: Bayes factor. Column 3 and 4: information criteria AIC and BIC, respectively.}
\label{table:evidence1}
\end{table}

\begin{table}[]
\centering
\begin{tabular}{|c |c |c |c |c |c |c|} 
\hline
\bf{Model}& $\chi^2$& $\bf{q_0}$ & $\bf{j_0}$ & \bf{$\ln{B_{ij}}$}   & \bf{AIC}   & \bf{BIC}  \\
\hline\hline
$\Lambda$CDM      &    7.991       &         $-0.550 \pm  0.020$                                   &      $1.0$       &       -     &            -    &     -          \\
\hline
CPL           &        2.75       &            $-1.721 \pm  6.564$                               &         $11.716\pm 66.16$    &   2.163         &     10.72               & 30.659           \\
\hline
RS            &       3.81        &        $-0.632 \pm  0.8272$                                    &    $0.344  \pm 2.7626$        &     2.281       &      11.81                 &   31.749     \\
\hline
$f(z)$CDM Pad\'e-like      &    7.97         &    $ -0.910\pm 0.142$     & $2.00 \pm 0.601$           & 7.594           &   15.97       & 35.909 \\    
\hline
$f(z)$CDM Chebyshev-like      &    7.96         &    $4.358\pm 5.989$         & $-7.643\pm 12.606$           & 7.588            &  17.96        & 42.884 \\                  
\hline
\end{tabular}
\caption{Cosmological models and approximants using trained DL Pantheon sample. Column 1: cosmographic parameters values at 2-$\sigma$. Column 2: Bayes factor. Column 3 and 4: information criteria AIC and BIC, respectively.}
\label{table:evidence2}
\end{table}

\begin{table}[]
\centering
\begin{tabular}{|c |c |c |c |c |c |c|} 
\hline
\bf{Model}& $\chi^2$& $\bf{q_0}$ & $\bf{j_0}$ & \bf{$\ln{B_{ij}}$}   & \bf{AIC}   & \bf{BIC}  \\
\hline\hline
$\Lambda$CDM      &       1035    &          $-0.551\pm 0.016$                                 &     $1.0$       &     -       &       -         &     -          \\
\hline
CPL           &    1035           &       $-0.475  \pm   0.380$  &      $0.224\pm 2.240$      &    2.637        &    1062.94                &       1043    \\
\hline
RS            &        1035       &      $ -0.505 \pm     0.286$                              &    $0.634\pm1.309$       &     2.637      &        1062.94              & 1043      \\
\hline
$f(z)$CDM Pad\'e-like      & 1035            &    $-0.552 \pm 0.35$                           & $1.070\pm 0.281$            &  2.637          &  1043        & 1062.94 \\    
\hline
$f(z)$CDM Chebyshev-like      &  1035           &    $0.525\pm 0.445$               & $0.947\pm 0.605$            &   3.042         &  1045        & 1069.92 \\                  
\hline
\end{tabular}
\caption{Cosmological models and approximants using Pantheon + DL Pantheon samples. Column 1: cosmographic parameters values at 2-$\sigma$. Column 2: Bayes factor. Column 3 and 4: information criteria AIC and BIC, respectively.}
\label{table:evidence3}
\end{table}

\begin{figure*}
    \centering
    \includegraphics[width=0.67\textwidth,origin=c,angle=0]{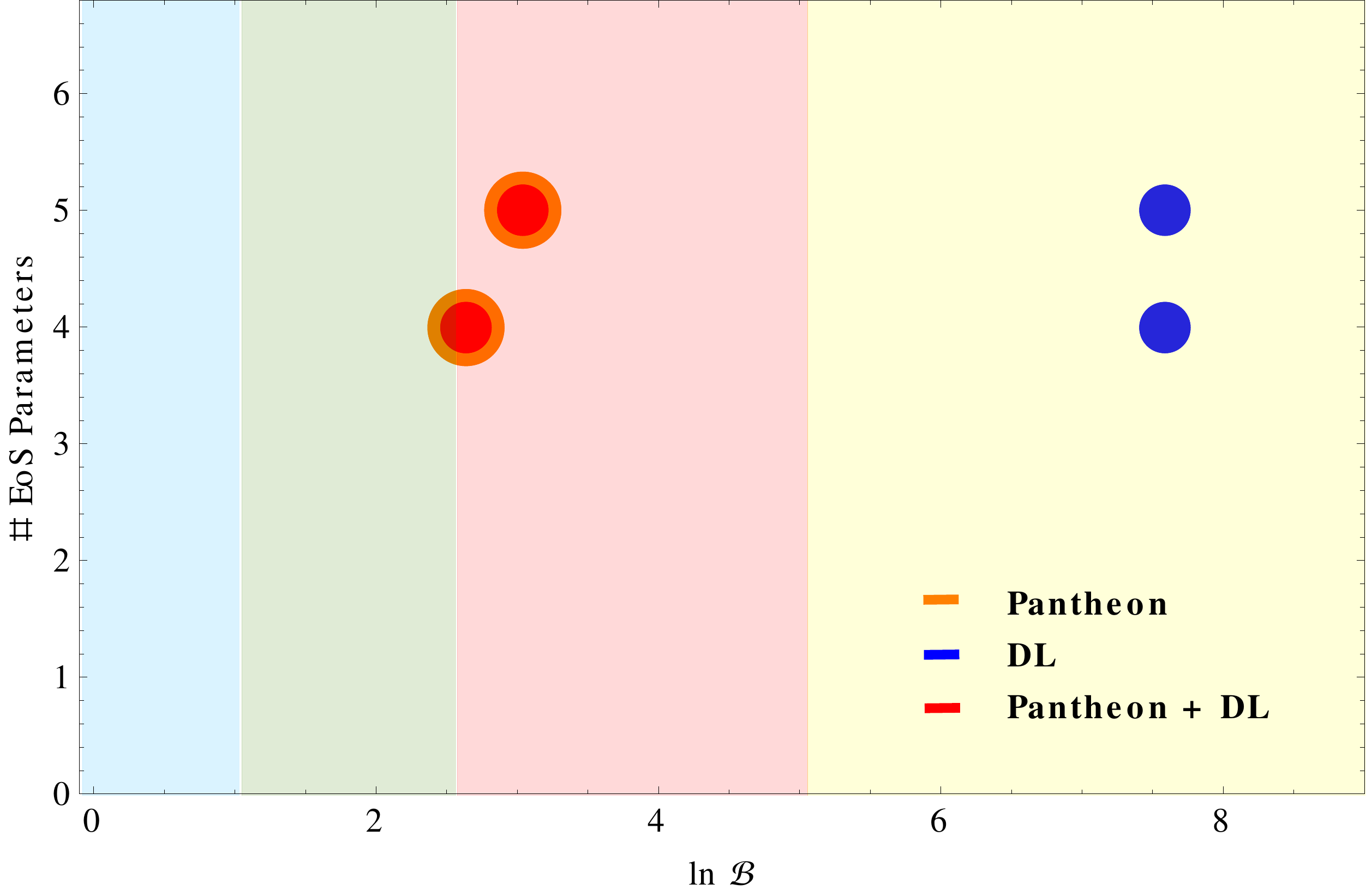}
          \caption{Bayesian portrait for Pad\'e (2,2) and Chebyshev (2,1) according to the number of free EoS parameters using the values given in Tables \ref{tab:pade_observational}-\ref{tab:pade_DL}-\ref{tab:pade_DL_obs} and \ref{tab:chebyshev_observational}-\ref{tab:chebyshev_DL}-\ref{tab:chebyshev_DL_obs}. The qualitative color regions represent Jeffrey's scale as: strong evidence in favour of hypothesis model (light-yellow), moderate evidence in favour of hypothesis model (light-red), weak evidence for hypothesis model (light-green) and inconclusive evidence (light-blue).
          Each color in the plot legend indicates the sample used
to calculated the value $\ln B_{ij}$ in comparison to $\Lambda$CDM model. 
\textbf{The Pad\'e model lies in the y-axes line corresponding to four free parameters and the Chebyshev model lies in the y-axes line corresponding to five free parameters. Notice that the results obtained by using Pantheon and Pantheon+DL samples overlap statistically. }
}
    \label{fig:bayesian_comp2}
\end{figure*}

\begin{figure*}
\centering
\includegraphics[width=0.68\textwidth,origin=c,angle=0]{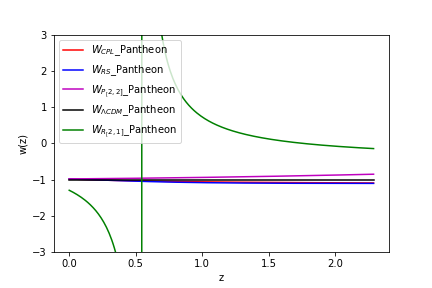}
\includegraphics[width=0.68\textwidth,origin=c,angle=0]{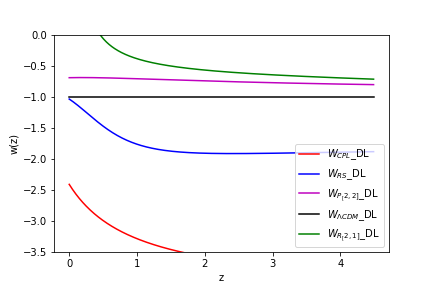}
\includegraphics[width=0.68\textwidth,origin=c,angle=0]{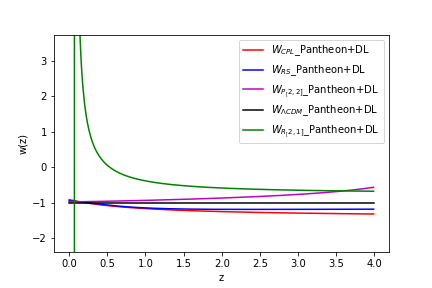}
\caption{Equation of state for the standard cosmological models Sec.(\ref{ssec:eos_standards}) and $f(z)$CDM cosmological models Secs.(\ref{ssec:pade}-\ref{ssec:chebyshev}) obtained using the best fits reported for each sample and cases. $\Lambda$CDM model is denote by a solid black line, the standard models CPL and RS by solid red and blue lines, respectively and the Pad\'e (2,2) and  Chebyshev (2,1) models by purple and green lines, respectively.} 
\label{fig:eos_tested}
\end{figure*}


\section{Discussion}
\label{sec:discussion}
In this work, we presented a proposal that consists of an \textit{inverse cosmography} approach from which we can compute a \textit{generic} EoS without assuming directly a cosmographic series. In this approach it is possible to relax a couple of cosmography problems, as e.g truncation issues over the series/polynomials proposed. Also, we developed the study of the dark energy dynamics, without a $\Lambda$, by proposing suitable Pad\'e or Chebyshev polynomials evolving with a cold dark matter (CDM) fluid. All these models were tested with current supernovae data and deep learning for supernovae trained samples. This machine learning architecture can train a homogeneous SNeIa sample in where the resulting dynamics of dark energy could lead to the necessity of another cosmological model different from $\Lambda$CDM. In this panorama, our inverse cosmography approach can fit very well since it
is not necessary to consider higher-order corrections of the cosmography series to obtain a convergence best fit in comparison to the standard dark energy EoSs as our Bayesian analyses show (see Tables \ref{table:evidence1}-\ref{table:evidence2}-\ref{table:evidence3}). The advantage to join current and trained supernovae sample is the density information data that we acquire in a redshift range of $0.01 < z < 4$, which is larger than the observable one. From the results of the Bayesian evidence, we notice that models polynomial-like, as Pad\'e and Chebyshev lie in better significant regions in comparison to the standard ones  (see Figure \ref{fig:bayesian_comp}). Furthermore, $f(z)$CDM Pad\'e-like and $f(z)$CDM Chebyshev-like are decisive in comparison to $\Lambda$CDM using DL sample (see Figure \ref{fig:bayesian_comp2}).

Now, from the perspective of the EoS's (see Figure \ref{fig:eos_tested}), we obtained the following results: in the case $f(z)$CDM Pad\'e-like we obtain a dynamical EoS with a transition from $\Lambda$CDM to a quintessence scenario. A deviation of 1-$\sigma$ from the standard model can be observed using Pantheon and join sample. For this case, we notice that $f(z)$CDM Pad\'e-like model does not have any divergence in the observational redshift range as it was reported in the mentioned references. Also, this model mimics $\Lambda$CDM at low redshift. On the contrary, $f(z)$CDM Chebyshev-like model does present a divergence at low redshift, but at high redshift starts to mimic $f(z)$CDM Pad\'e-like model. Notice how all the tested models reproduce the theoretical scenarios reported in Figure \ref{fig:eos_theoretical}.

Furthermore, up to the best of our knowledge, no CLASS + MontePython code adapted for cosmography has yet been developed. This process involves introducing directly the definition of the modulus distance and then deal with series divergencies. Meanwhile, using inverse cosmography it is possible to adapt easily the numerical codes and obtain a theoretical convergence\footnote{The code will be released after publication.}. The inclusion of observable data from another nature that can give us information a higher redshifted universe will enrich the convergence of a series with more exotic polynomials. This future study will be reported elsewhere.


\acknowledgments
CE-R acknowledges the \textit{Royal Astronomical Society} as FRAS 10147. The authors are supported by \textit{PAPIIT} Project IA100220. This article is also based upon work from COST action CA18108, supported by COST (European Cooperation in Science and Technology). The Authors also thanks M. Carvajal, H. Quevedo, O. Luongo for their useful discussions. CE-R thanks J. Levi Said for his insights ideas.

\begin{appendix}
\label{sec:appendix}

\section{Generic cosmographic parameters}

In this Appendix we present the exact solutions of the cosmographic parameters for the models considered and in term of the standard cosmological parameters.

\subsection{Chevallier-Polarski-Linder (CPL) case}

With the inverse cosmography prescription  (\ref{eq:q})-(\ref{eq:j})-(\ref{eq:s}), we can obtain the following cosmographic parameters for this model as follow:
\begin{eqnarray}
q(z) &=&\frac{\left(\Omega _m-1\right) (z+1)^{3 \left(w_a+w_0\right)} \left(3 z w_a+3 w_0 (z+1)+z+1\right)-(z+1)
   \Omega _m e^{\frac{3 z w_a}{z+1}}}{2 \left(\Omega _m-1\right) (z+1)^{3 w_a+3 w_0+1}-2 (z+1) \Omega _m
   e^{\frac{3 z w_a}{z+1}}}, \\
   j(z)&=& \frac{\left(\Omega _m-1\right) \left(9 z^2 w_a^2+3 (3 z+1) (z+1) w_a+2 (z+1)^2\right) (z+1)^{3
   \left(w_a+w_0\right)}}{2
   (z+1)^2 \left(\left(\Omega _m-1\right) (z+1)^{3 \left(w_a+w_0\right)}-\Omega _m e^{\frac{3 z
   w_a}{z+1}}\right)} \nonumber\\ && 
   +\frac{9 w_0 \left(\Omega _m-1\right) \left(2 z w_a+z+1\right) (z+1)^{3 w_a+3 w_0+1}}{2
   (z+1)^2 \left(\left(\Omega _m-1\right) (z+1)^{3 \left(w_a+w_0\right)}-\Omega _m e^{\frac{3 z
   w_a}{z+1}}\right)} \nonumber\\ &&
   +\frac{9 w_0^2 \left(\Omega _m-1\right) (z+1)^{3 w_a+3 w_0+2}-2 (z+1)^2 \Omega _m e^{\frac{3 z w_a}{z+1}}}{2
   (z+1)^2 \left(\left(\Omega _m-1\right) (z+1)^{3 \left(w_a+w_0\right)}-\Omega _m e^{\frac{3 z
   w_a}{z+1}}\right)},\\   
   s(z) &=& \left[\frac{\left(\Omega _m-1\right) (z+1)^{3 \left(w_a+w_0\right)} }{2 \left(\Omega _m-1\right) (z+1)^{3 w_a+3 w_0+4}-2 (z+1)^4 \Omega _m
   e^{\frac{3 z w_a}{z+1}}}  \right]\nonumber\\ &&
   \times
   (-9 z^3 w_a^3+(z+1)^2 w_a \left(-9 w_0
   \left(3 w_0 z-2 (z-1) z+1\right)+9 z^2+z-2\right)
   \nonumber\\&&
   +9 z (z+1) w_a^2 \left(-3 w_0 z+(z-1)
   z-1\right)
 \nonumber\\&&
   +\left(9 w_0 \left(w_0+1\right)+2\right) (z+1)^3 \left(z-w_0\right))-2 z (z+1)^3 \Omega
   _m e^{\frac{3 z w_a}{z+1}},
\end{eqnarray}

\subsection{Redshift squared (RS) case}

With the inverse cosmography prescription  (\ref{eq:q})-(\ref{eq:j})-(\ref{eq:s}), we can obtain the following cosmographic parameters for this model as follow:

\begin{eqnarray}
\frac{2}{3}(q(z) +1) &=&\frac{\left(\Omega _m-\left(\Omega _m-1\right) (z+1)^{3 w_0}
   \left(z^2+1\right)^{\frac{3 w_a}{2}-1} \left(z (z+1) w_a+\left(w_0+1\right)
   \left(z^2+1\right)\right)\right)}{\Omega _m-\left(\Omega _m-1\right) (z+1)^{3 w_0}
   \left(z^2+1\right)^{\frac{3 w_a}{2}}}, \quad\quad \\
   2 j(z)&=& \frac{3 \left(\Omega _m-1\right) (z+1)^{3 w_0}} {\left(z^2+1\right)^2 \left(\left(\Omega _m-1\right) (z+1)^{3 w_0}-\Omega _m
   \left(z^2+1\right)^{-\frac{3 w_a}{2}}\right)}
      \nonumber\\&&
   \times
   ((z+1) w_a \left(6 w_0
   \left(z^3+z\right)+z (z (3 z-1)+5)+1\right)
      \nonumber\\&&
   +3 z^2 (z+1)^2 w_a^2+3 w_0 \left(w_0+1\right)
   \left(z^2+1\right)^2)
  +2,\\    
   s(z) &=& \frac{3 \left(\Omega _m-1\right) (z+1)^{3 w_0}}{6 (z+1) \left(z^2+1\right)^3
   \left(\left(\Omega _m-1\right) (z+1)^{3 w_0} \left(z^2+1\right)^{\frac{3 w_a}{2}}-\Omega _m\right)} 
   \nonumber\\&&
   \times
   \left(z^2+1\right)^{\frac{3 w_a}{2}} (-9 z^3 (z+1)^3
   w_a^3+9 z (z+1)^2 w_a^2 (-3 w_0 \left(z^3+z\right)
    \nonumber\\&&
   +z (z ((z-1) z+2)-3)-1)+(z+1) w_a (9
   w_0 \left(z^2+1\right) 
    \nonumber\\&&
    \times
    \left(-3 w_0 \left(z^3+z\right)+z (z (2 (z-1) z+3)-4)-1\right)
      \nonumber\\&&
    +z (z (z (z (z (9
   z-5)+26)-8)+27)-3)-6)
     \nonumber\\&&
   +\left(9 w_0 \left(w_0+1\right)+2\right) \left(z^2+1\right)^3
   \left(z-w_0\right))-6 z \left(z^2+1\right)^3 \Omega _m.
\end{eqnarray}

\subsection{$f(z)$CDM Pad\'e-like case}
With the inverse cosmography prescription  (\ref{eq:q})-(\ref{eq:j})-(\ref{eq:s}), we can obtain the following cosmographic parameters for this model as follow:

\begin{eqnarray}
q(z) &=&\frac{(z+1) \left(3 (z+1)^2 \Omega _m+\frac{\Omega _p \left(P_1 \left(Q_2 z^2-1\right)+P_2 z \left(Q_1
   z+2\right)-2 Q_2 z-Q_1\right)}{\left(z \left(Q_2 z+Q_1\right)+1\right){}^2}\right)}{2 \left((z+1)^3
   \Omega _m+\frac{\Omega _p \left(P_2 z^2-P_1 z+1\right)}{z \left(Q_2 z+Q_1\right)+1}\right)}-1, 
   \end{eqnarray}
   
   \begin{eqnarray}
  \frac{ j(z)}{(z+1) \Omega _p} -1&=& \frac{P_1 \left(Q_2^2 (-(2 z+1)) z^3+Q_1 \left(-Q_2 z^3+2 z+1\right)+3 Q_2 (z+1)
   z+1\right)}  {\left(z \left(Q_2 z+Q_1\right)+1\right){}^2 \left((z+1)^3 \Omega _m \left(z
   \left(Q_2 z+Q_1\right)+1\right)+\Omega _p \left(P_2 z^2-P_1 z+1\right)\right)}
    \nonumber\\&&
   +\frac{P_2 \left(z^2 \left(-Q_1 \left(Q_1 z+3\right)-Q_2 \left(Q_1 (2 z+1) z+5
   z+3\right)\right)-z+1\right)}{\left(z \left(Q_2 z+Q_1\right)+1\right){}^2 \left((z+1)^3 \Omega _m \left(z
   \left(Q_2 z+Q_1\right)+1\right)+\Omega _p \left(P_2 z^2-P_1 z+1\right)\right)}
      \nonumber\\&&
   +\frac{Q_2^2 z^2 (5 z+3)+Q_2 \left(3 Q_1 (2 z+1) z+z-1\right)+Q_1 \left(2 Q_1
   z+Q_1+1\right))}
   {\left(z \left(Q_2 z+Q_1\right)+1\right){}^2 \left((z+1)^3 \Omega _m \left(z
   \left(Q_2 z+Q_1\right)+1\right)+\Omega _p \left(P_2 z^2-P_1 z+1\right)\right)}, \quad \quad \quad 
   \end{eqnarray}
   
   \begin{eqnarray}
   \frac{1-s(z)}{(z+1)}&=&\frac{(z+1) \Omega _m \left(z \left(Q_2 z+Q_1\right)+1\right){}^4}  {\left(z \left(Q_2 z+Q_1\right)+1\right){}^3 \left((z+1)^3 \Omega _m \left(z \left(Q_2
   z+Q_1\right)+1\right)+\Omega _p \left(P_2 z^2-P_1 z+1\right)\right)}
      \nonumber\\&&
   +\frac{\Omega _p (Q_2^3
   z^3 \left(P_1 z (2 z (z+1)+1)
   -z (5 z+7)-4\right)} {\left(z \left(Q_2 z+Q_1\right)+1\right){}^3 \left((z+1)^3 \Omega _m \left(z \left(Q_2
   z+Q_1\right)+1\right)+\Omega _p \left(P_2 z^2-P_1 z+1\right)\right)}
       \nonumber\\&&
   +\frac{Q_2^2 z \left(P_1 z \left(Q_1 (3 z+1) z^2-(z+8)
   z-6\right)-Q_1 (z (11 z+12)+6) z-6 z^2+2 z+4\right)} {\left(z \left(Q_2 z+Q_1\right)+1\right){}^3 \left((z+1)^3 \Omega _m \left(z \left(Q_2
   z+Q_1\right)+1\right)+\Omega _p \left(P_2 z^2-P_1 z+1\right)\right)}
        \nonumber\\&&
   +\frac{Q_2 (P_1 \left(Q_1^2 z^4-4 Q_1 (z+1)^2 z-2
   (2 z+1) z+1\right)}  {\left(z \left(Q_2 z+Q_1\right)+1\right){}^3 \left((z+1)^3 \Omega _m \left(z \left(Q_2
   z+Q_1\right)+1\right)+\Omega _p \left(P_2 z^2-P_1 z+1\right)\right)}
       \nonumber\\&&
   +\frac{2 Q_1 \left(-2 Q_1 (2 z (z+1)+1) z-4 z^2+1\right)-z)+P_2 (Q_2^2 z^3
   (Q_1 z (2 z (z+1)+1)} {\left(z \left(Q_2 z+Q_1\right)+1\right){}^3 \left((z+1)^3 \Omega _m \left(z \left(Q_2
   z+Q_1\right)+1\right)+\Omega _p \left(P_2 z^2-P_1 z+1\right)\right)}
     \nonumber\\&&
   +\frac{z (5 z+7)+4)+Q_2 z \left(Q_1 z^2 \left(Q_1 (3 z+1) z+10
   z+4\right)+2 (z-1) (3 z+2)\right)}  {\left(z \left(Q_2 z+Q_1\right)+1\right){}^3 \left((z+1)^3 \Omega _m \left(z \left(Q_2
   z+Q_1\right)+1\right)+\Omega _p \left(P_2 z^2-P_1 z+1\right)\right)}
     \nonumber\\&&
   +\frac{Q_1 \left(Q_1 z^3 \left(Q_1 z+4\right)+4 z^2-2
   z-1\right)+z)}  {\left(z \left(Q_2 z+Q_1\right)+1\right){}^3 \left((z+1)^3 \Omega _m \left(z \left(Q_2
   z+Q_1\right)+1\right)+\Omega _p \left(P_2 z^2-P_1 z+1\right)\right)}
     \nonumber\\&&
   +\frac{Q_1 \left(-\left(P_1+Q_1\right) \left(Q_1 (2 z (z+1)+1)+3
   z\right)-Q_1\right)}  {\left(z \left(Q_2 z+Q_1\right)+1\right){}^3 \left((z+1)^3 \Omega _m \left(z \left(Q_2
   z+Q_1\right)+1\right)+\Omega _p \left(P_2 z^2-P_1 z+1\right)\right)}
     \nonumber\\&&
   -\frac{\left(P_1+1\right) Q_1+Q_2)-\left(P_1+P_2\right) \Omega
   _p}
   {\left(z \left(Q_2 z+Q_1\right)+1\right){}^3 \left((z+1)^3 \Omega _m \left(z \left(Q_2
   z+Q_1\right)+1\right)+\Omega _p \left(P_2 z^2-P_1 z+1\right)\right)}, \quad \quad \quad \quad 
\end{eqnarray}

\subsection{$f(z)$CDM Chebyshev-like case}
With the inverse cosmography prescription  (\ref{eq:q})-(\ref{eq:j})-(\ref{eq:s}), we can obtain the following cosmographic parameters for this model as follow:

\begin{eqnarray}
q(z) &=& \frac{(z+1) \left(\Omega _p \left(2 a_2 z \left(b_1 z+2\right)-a_3 b_1+a_1\right)+3 (z+1)^2 \left(b_1
   z+1\right){}^2 \Omega _m\right)}{2 \left(b_1 z+1\right) \left(\left(z \left(2 a_2
   z+a_1\right)+a_3\right) \Omega _p+(z+1)^3 \left(b_1 z+1\right) \Omega _m\right)}-1, \\
j(z)&=& \frac{ \Omega _p(a_1 \left(b_1 \left(b_1 z^3-3 z-1\right)-1\right)+b_1(a_3 \left(b_1 (3 z
   (z+1)+1)+3 z+1\right)} {\left(b_1 z+1\right){}^2 \left(\left(z \left(2 a_2
   z+a_1\right)+a_3\right) \Omega _p+(z+1)^3 \left(b_1 z+1\right) \Omega _m\right)}
     \nonumber\\&&
   -\frac{2 a_2 z^2 \left(b_1 z+z+3\right))+2 a_2+a_3)+(z+1)^3 \left(b_1
   z+1\right){}^3 \Omega _m}
   {\left(b_1 z+1\right){}^2 \left(\left(z \left(2 a_2
   z+a_1\right)+a_3\right) \Omega _p+(z+1)^3 \left(b_1 z+1\right) \Omega _m\right)},\\
   s(z) &=&\frac{\Omega _p (a_1 \left(b_1 \left(b_1 z^3-3 z-1\right)-1\right)+b_1 (a_3 \left(b_1 (3 z
   (z+1)+1)+3 z+1\right)} {\left(b_1 z+1\right){}^2 \left(\left(z \left(2 a_2
   z+a_1\right)+a_3\right) \Omega _p+(z+1)^3 \left(b_1 z+1\right) \Omega _m\right)}
     \nonumber\\&&
   -\frac{2 a_2 z^2 \left(b_1 z+z+3\right))+2 a_2+a_3)+(z+1)^3 \left(b_1
   z+1\right){}^3 \Omega _m}
   {\left(b_1 z+1\right){}^2 \left(\left(z \left(2 a_2
   z+a_1\right)+a_3\right) \Omega _p+(z+1)^3 \left(b_1 z+1\right) \Omega _m\right)}.
\end{eqnarray}

\end{appendix}


\begin{thebibliography}{0} 

\bibitem{Verde:2019ivm} 
  L.~Verde, T.~Treu and A.~G.~Riess,
  Nature Astronomy 2019
  doi:10.1038/s41550-019-0902-0
  [arXiv:1907.10625 [astro-ph.CO]].
  
  \bibitem{Bolotin:2018xtq} 
  Y.~L.~Bolotin, V.~A.~Cherkaskiy, O.~Y.~Ivashtenko, M.~I.~Konchatnyi and L.~G.~Zazunov,
  arXiv:1812.02394 [gr-qc].
  
    \bibitem{Capozziello:2013wha} 
  S.~Capozziello, M.~De Laurentis, O.~Luongo and A.~Ruggeri,
  Galaxies {\bf 1}, 216 (2013)
  doi:10.3390/galaxies1030216
  [arXiv:1312.1825 [gr-qc]].
  
  \bibitem{Escamilla-Rivera:2019aol} 
  C.~Escamilla-Rivera and S.~Capozziello,
  Int.\ J.\ Mod.\ Phys.\ D {\bf 28}, no. 12, 1950154 (2019)
  doi:10.1142/S0218271819501542
  [arXiv:1905.04602 [gr-qc]].

\bibitem{Escamilla-Rivera:2019hqt}
C.~Escamilla-Rivera, M.~A.~C.~Quintero and S.~Capozziello,
JCAP \textbf{03} (2020) no.03, 008
doi:10.1088/1475-7516/2020/03/008
[arXiv:1910.02788 [astro-ph.CO]].
  
  \bibitem{Capozziello:2019cav} 
  S.~Capozziello, R.~D'Agostino and O.~Luongo,
  Int.\ J.\ Mod.\ Phys.\ D {\bf 28}, no. 10, 1930016 (2019)
  doi:10.1142/S0218271819300167
  [arXiv:1904.01427 [gr-qc]].
  
  \bibitem{spec_surveys}
  \url{https://www.desi.lbl.gov}
  \url{https://www.lsst.org} 
  
  \bibitem{Schlegel:2019eqc} 
  D.~J.~Schlegel {\it et al.},
  arXiv:1907.11171 [astro-ph.IM].
 
  \bibitem{EDGES}
 Bowman, J., Rogers, A., Monsalve, R. et al. 
 Nature 555, 67–70 (2018). 
 
\bibitem{Corman:2020pyr}
M.~Corman, C.~Escamilla-Rivera and M.~Hendry,
[arXiv:2004.04009 [astro-ph.CO]].
 
\bibitem{LIGOScientific:2020stg}
 [LIGO Scientific and Virgo],
[arXiv:2004.08342 [astro-ph.HE]].

  \bibitem{Escamilla-Rivera:2015odt} 
  C.~Escamilla-Rivera and J.~C.~Fabris,
  Galaxies {\bf 4}, no. 4, 76 (2016)
  doi:10.3390/galaxies4040076
  [arXiv:1511.07066 [astro-ph.CO]].
 
  
  \bibitem{Li:2019qic} 
  E.~K.~Li, M.~Du and L.~Xu,
  Mon.\ Not.\ Roy.\ Astron.\ Soc.\  {\bf 491}, no. 4, 4960 (2020)
  doi:10.1093/mnras/stz3308
  
  \bibitem{Gruber:2013wua} 
  C.~Gruber and O.~Luongo,
  Phys.\ Rev.\ D {\bf 89}, no. 10, 103506 (2014)
  doi:10.1103/PhysRevD.89.103506
  
  \bibitem{Capozziello:2017nbu} 
  S.~Capozziello, R.~D'Agostino and O.~Luongo,
  Mon.\ Not.\ Roy.\ Astron.\ Soc.\  {\bf 476}, no. 3, 3924 (2018)
  doi:10.1093/mnras/sty422
  [arXiv:1712.04380 [astro-ph.CO]].
  
  \bibitem{Aviles:2012ay} 
  A.~Aviles, C.~Gruber, O.~Luongo and H.~Quevedo,
  Phys.\ Rev.\ D {\bf 86}, 123516 (2012)
  doi:10.1103/PhysRevD.86.123516
  [arXiv:1204.2007 [astro-ph.CO]].
  
  \bibitem{Benetti:2019gmo} 
  M.~Benetti and S.~Capozziello,
  JCAP {\bf 1912}, no. 12, 008 (2019)
  doi:10.1088/1475-7516/2019/12/008
  [arXiv:1910.09975 [astro-ph.CO]].
  
 \bibitem{Escamilla-Rivera:2016qwv} 
  C.~Escamilla-Rivera,
  Galaxies {\bf 4} (2016) 8. 
\bibitem{Chevallier:2000qy}
  M. Chevallier,  D. Polarski, 
   Int.\ J.\ Mod.\ Phys.\ D {\bf 10} (2001) 213.

  
  \bibitem{Linder:2007wa}
 E.V.  Linder, 
   Gen.\ Rel.\ Grav. {\bf  40 } (2008) 329.

  
  \bibitem{Barboza:2008rh}
 E.M. Jr. Barboza,  J.S. Alcaniz, 
    Phys.\ Lett.\ B  {\bf 666}  (2008)  415.


\bibitem{Aviles:2016wel} 
  A.~Aviles, J.~Klapp and O.~Luongo,
  Phys.\ Dark Univ.\  {\bf 17}, 25 (2017)
  doi:10.1016/j.dark.2017.07.002
  [arXiv:1606.09195 [astro-ph.CO]].
  
  \bibitem{Capozziello:2018jya} 
  S.~Capozziello, Ruchika and A.~A.~Sen,
  Mon.\ Not.\ Roy.\ Astron.\ Soc.\  {\bf 484}, 4484 (2019)
  doi:10.1093/mnras/stz176
  [arXiv:1806.03943 [astro-ph.CO]].
  
\bibitem{Vitagliano:2009et}
V.~Vitagliano, J.~Q.~Xia, S.~Liberati and M.~Viel,
JCAP \textbf{03}, 005 (2010)
doi:10.1088/1475-7516/2010/03/005
[arXiv:0911.1249 [astro-ph.CO]].

\bibitem{Liddle:2004nh} 
  A.~R.~Liddle,
  Mon.\ Not.\ Roy.\ Astron.\ Soc.\  {\bf 351}, L49 (2004)
  doi:10.1111/j.1365-2966.2004.08033.x
  [astro-ph/0401198].

    \bibitem{jeffreys}
 H. Jeffreys,  \emph{Theory of Probability}, 3rd ed.; Oxford University Press: Oxford, United Kingdom. 1998.
  
  \bibitem{Scolnic:2017caz} 
  D.~M.~Scolnic {\it et al.},
  Astrophys.\ J.\  {\bf 859}  (2018) 101



\end{thebibliography}
\end{document}